\documentclass[compsoc,conference,a4paper,10pt,]{IEEEtran}%
\pdfoutput=1
\IEEEoverridecommandlockouts

\usepackage{cite}
\usepackage{amsmath,amssymb,amsfonts}
\usepackage{algorithmic}
\usepackage{graphicx}
\usepackage{textcomp}
\usepackage{bmpsize}
\usepackage{xcolor}
\usepackage{lipsum}
\usepackage[colorlinks=true,urlcolor=black]{hyperref}
\def\BibTeX{{\rm B\kern-.05em{\sc i\kern-.025em b}\kern-.08em
    T\kern-.1667em\lower.7ex\hbox{E}\kern-.125emX}}

\usepackage{colortbl} %
\usepackage{array} %
\newcolumntype{P}[1]{>{\centering\arraybackslash}p{#1}} %
\usepackage{pifont} %
\usepackage{adjustbox} %
\usepackage{booktabs} %
\usepackage{makecell} %

\usepackage{tikz}
\newcommand*\circled[1]{\tikz[baseline=(char.base)]{
            \node[shape=circle,draw,inner sep=0.8pt] (char) {#1};}}

\def\LgenTspec{\textbf{\circled{1}}}
\def\TgenLspec{\textbf{\circled{2}}}
\def\EitherLorT{\textbf{\circled{3}}}
\def\NotinDash{\textbf{\circled{4}}}
\def\TrafnotinL{\textbf{\circled{5}}}

\newif\ifediting
\editingtrue  %
\definecolor{mygray}{gray}{0.2}
\definecolor{mylightgray}{gray}{0.1}
\definecolor{mycyan}{rgb}{0.05, 0.6, 0.85}
\definecolor{mygreen}{rgb}{0.00, 0.67, 0.51}

\ifediting
    \usepackage{comment}
    \usepackage[draft,textsize=footnotesize,textwidth=17mm]{todonotes}
\fi

\usepackage{amsthm}

\theoremstyle{definition}

\theoremstyle{definition}
\newtheorem{potential-violation}{Potential violation}

\theoremstyle{definition}
\newtheorem{legal-issue}{Legal issue}

\theoremstyle{definition}

\def\adstorage{\texttt{ad\_storage} }
\def\analyticsstorage{\texttt{analytics\_storage} }
\def\aduserdata{\texttt{ad\_user\_data} }
\def\adpersonalisation{\texttt{ad\_personalisation} }

\def\code#1{\texttt{#1}}
\def\gtmjs{\code{gtm.js} }

\def\tagstotal{718} %

\newcommand{\linebreakand}{%
  \end{@IEEEauthorhalign}
  \hfill\mbox{}\par
  \mbox{}\hfill\begin{@IEEEauthorhalign}
}

\begin{document}

\title{You Can’t Trust Your Tag Neither: Privacy Leaks and Potential Legal Violations within the Google Tag Manager}

\author{
\IEEEauthorblockN{Gilles Mertens}
\IEEEauthorblockA{
 Inria Centre at University \\
Grenoble-Alpes, France \\
gilles.mertens@inria.fr}
\and
\IEEEauthorblockN{Nataliia Bielova}
\IEEEauthorblockA{
 Inria Centre at University \\
Côte d'Azur, France \\
nataliia.bielova@inria.fr}
\and 
\IEEEauthorblockN{Vincent Roca}
\IEEEauthorblockA{
 Inria Centre at University \\
Grenoble-Alpes, France \\
vincent.roca@inria.fr}
\and
\IEEEauthorblockN{Cristiana Santos}
\IEEEauthorblockA{
Utrecht University \\
The Netherlands \\
c.teixeirasantos@uu.nl}
}

\maketitle

\begin{abstract}
Tag Management Systems (TMS) were developed in order to support website Publishers in installing multiple third-party JavaScript scripts (Tags) on their websites. 
Google has proposed its own TMS called ``Google Tag Manager'' (GTM) that is currently present on 52\% of the top 1 million most popular websites.
However, GTM has not yet been thoroughly evaluated by the academic research community. 
In this work, we study, for the first time, the Tags provided within the GTM system.
Our methodology consists in installing Tags in isolation to analyze the types of data that Tags collect and contrast them to the legal and technical documentation, in collaboration with a legal expert. 
Across three studies -- in-depth analysis of 6 Tags, automated analysis of \tagstotal\ Tags, and analysis of Google ``Consent Mode'' -- we discover multiple hidden data leaks, incomplete and diverging declarations, undisclosed third-parties and cookies, personal data sharing without consent and we further identify potential legal violations within EU Data Protection law. 

\end{abstract}


\section{Introduction} 
\label{sec:intro}

Over the last decade, researchers have demonstrated that third-parties collect users' data with the help of third-party JavaScript scripts~\cite{nikiforakis_you_2012}. 
Since early 2010, the research field of detection and large-scale measurement of  third-party Web tracking has evolved, measuring both stateful~\cite{Roes-etal-12-NSDI,Lern-etal-16-USENIX,Engl-etal-16-CCS,Bash-etal-16-USENIX,Yu-etal-16-WWW,Bash-Wils-18-PETS,Iord-etal-18-IMC,Foua-etal-20-PoPETs} and 
stateless tracking~\cite{Acar-etal-13-CCS,Acar-etal-14-CCS,Lape-etal-20-TWEB,Bahr-etal-22-PETS,Foua-etal-22-PETs}. At the same time, policy-makers worldwide have issued new laws and regulations to protect users from such massive data collection by third-party companies. In the EU, the ePrivacy Directive (ePD)~\cite{ePD-09} and  the General Data Protection Regulation (GDPR) require user consent~\cite{GDPR,ePD-09,Sant-etal-20-TechReg}. 
These laws have impacted the Web privacy measurement research community, resulting in a new area of automated compliance measurement that has since emerged~\cite{Matt-etal-20-IEEESP,Sanc-etal-19-AsiaCCS,Dege-etal-19-NDSS,Urba-etal-20-AsiaCCS,Nouw-etal-20-CHI,Liu-etal-24-PoPETs}.

The main components of a website responsible for data collection are third-party JavaScript libraries, invisible to users, that are silently executed in the webpage's background  and are often called ``Tags'' in the Web marketing industry~\cite{google-tag-platform-overview}.
Initially, to install a Tag on a webpage, the website Publishers, who own a website, only needed to copy and paste an external JavaScript library reference to the webpage's source code. 
However, as Publishers were installing more and more Tags, %
manual Tag management became challenging and, as a result, ``Tag Management Systems'' (TMS) were developed by the industry.
TMS allow Publishers to install and configure Tags in a centralized manner without tinkering with the website source code. 
Once installed on a website, TMS automatically install and execute the third-party Tags selected by the Publisher. 

\begin{figure}[t]
        \centering
        \includegraphics[width=1\linewidth]{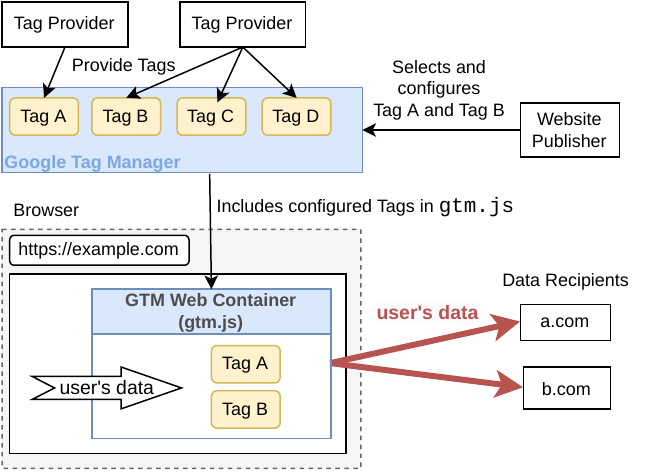}
        \caption{Client-side Google Tag Manager (GTM)}
        \label{fig:gtm-cs-arch} 
\end{figure}

Google developed its own TMS called ``Google Tag Manager'' (GTM) in 2012.
According to BuiltWith~\cite{gtm-prevalence-builtwith}, %
it is %
the most installed TMS on the market:
it is present on 52\% of the top 1 million websites and more than 60\% of the top 10K.
Despite its popularity, GTM has received very limited attention so far in the research community, and moreover, its compliance with the Data Protection laws, such as ePD and GDPR, has never been studied. 

Figure~\ref{fig:gtm-cs-arch} shows the GTM architecture proposed by Google that we call ``{Client-side GTM}''\footnote{%
Google also offers ``Server-side GTM'' that we do not study here.} since it loads %
Tags inside the user's browser within the ``GTM Web Container'' JavaScript library \gtmjs.
GTM is a free service for website Publishers. 
It offers a graphical interface and supports a seamless inclusion of %
marketing and analytic services implemented as Tags.
It is enough for a Publisher to select and properly configure desired Tags to include them in their websites. 
GTM  benefits from a community of contributors who create Tags for such services, whom we call ``Tag Providers''. 
Tags rely on the ability of browsers to communicate directly with third-party domains to %
set cookies and send users' data to %
third-parties, %
whom we call ``Data Recipients''. 

Differently from traditional large-scale measurement approaches, where all third-party services are analyzed together, in this paper, we %
study the behavior of Tags in isolation to %
\emph{detect %
privacy leaks in isolation}, allowing us to identify Tags that are responsible for sharing users' data further with other third-parties.
Our %
analysis method is inspired by previous research that analyzed the behavior of Consent Management Platforms (CMPs) in a fully controlled environment~\cite{toth_dark_2022,Sant-etal-21-APF} by installing them one-by-one %
and analyzing them in isolation. 
Our method is related to  the ``unitary asset testing methodology'' %
in software development, in which %
assets %
are configured and run one at a time, in a %
controlled execution environment to collect %
artifacts generated during its execution. 

Using this approach, we study Tags in a controlled environment, and we perform %
(1) an in-depth study of a limited number of most popular Tags, where we analyze the Tags' behavior and compare it to their documentation and dashboard; and %
(2) a large-scale study of  \tagstotal\  Tags, where we analyze network traffic to detect privacy leaks. 
Finally, (3) we study  a consent mechanism provided by Google -- ``Google Consent Mode'' -- to comply with the legal requirements for consent from the EU GDPR and ePrivacy Directive.
With this approach, we address the following questions:
\begin{itemize}
\item [\textbf{RQ1:}] What types of data do Tags collect, and do these  types include personal data according to EU law? %
\item [\textbf{RQ2:}] Do Data Recipients disclose what data is collected by the Tags they provide, and if so, do they collect it in a compliant way? %
\item [\textbf{RQ3:}] With whom Tags share users' data, and are companies  receiving such data declared by the Tag Providers, as required by the GDPR? %
\item [\textbf{RQ4:}] Does Google Tag collect personal data lawfully when  users reject consent under Consent Mode? %
\end{itemize}

To address these questions, we have set up a collaboration with a legal scholar in EU Data Protection law and we make the following contributions, where for each finding, we identified \emph{potential legal violations}:
\begin{itemize}
    
    \item We conduct an in-depth analysis of 6 popular Tags, and identify \emph{multiple inconsistencies} between the data disclosures in the documentation and data actually sent by these Tags in the network traffic.
    
    \item We perform \emph{the first large-scale automated analysis of \tagstotal\ Tags} in GTM, detecting that Tags not only set cookies and send users' data to their own servers without proper disclosures, but also include other third-parties with similar practices.

    \item We study ``Google Consent Mode'' in Google Tag and discover that data is \emph{actively collected via ``cookieless pings''} despite being declared as passively sent and despite user's consent refusal.

\end{itemize}

\section{Related Works and Background}
\label{sec:related-back}

GTM has become a popular tool for managing Tags on websites in recent years, 
however, to the best of our knowledge, no scientific study analyzed Tag Management Systems nor GTM in particular. 
This section introduces  related works, followed by  background knowledge on GTM and legal background on EU Data Protection laws that we rely upon in this paper. 

\subsection{Related Works}
\label{sec:related}

We discuss online tracking studies that measured the prevalence of Google services since GTM is served from a domain owned by Google, \texttt{googletagmanager.com}\cite{Foua-etal-24-PoPETs}.
We then review non-academic sources describing or analyzing GTM and research on privacy documentation.

\noindent
    \textbf{Online tracking and prevalence of Google.}
Multiple works have detected and measured third-party Web tracking at scale over the last decade,  
identifying Google's massive prevalence in the tracking ecosystem~\cite{Roes-etal-12-NSDI,Lern-etal-16-USENIX,Engl-etal-16-CCS,Bash-etal-16-USENIX,Yu-etal-16-WWW,Bash-Wils-18-PETS,Iord-etal-18-IMC,Foua-etal-20-PoPETs}. 
In 2009, Krishnamurthy and Wills~\cite{Krishnamurthy-etal-09-WWW} %
observed that requests to Google-owned servers happen on 60\% of the websites.
In 2016, Englehardt and Narayanan conducted an automated detection of stateful and stateless tracking on 1~million sites, showing that Google owned the five most prevalent third-party tracking domains, while \texttt{googletagmanager.com} was one of the top-20 third-party trackers%
~\cite[Section~5.1]{Engl-etal-16-CCS}. 
In 2020, Fouad et al.~\cite{Foua-etal-20-PoPETs} confirmed again the prevalence of Google -- they found the presence of tracking by Google domains on more than 85\% of the websites. 

\noindent
\textbf{Tag management and GTM.}
No academic research appears to have studied Tag management systems to date. 
However, several people close to AdTech circles share their experience and publish screenshots and other analyses of these tools on the web, either to help Publishers in their deployment or to point out legal or technical issues. 
The IT and digital marketing expert Julius Fedorovicius publishes courses and ebooks on GTM on his website \textit{Analytics Mania}~\cite{Analyticsmania-website}. 
The analytics developer Simo Ahava tests and comments on new functionalities in SEO and digital marketing tools. 
He maintains an extensive documentation on GTM~\cite{Ahava-GTM-documentation}, and in a recent article~\cite{Ahava-GTM-SST}, discusses GTM's limitations regarding the lack of control of the Publisher over the data collected by the Tags and problems regarding transparency for end-users.
The blogger \textit{Pixel de tracking}~\cite{PixeldeTracking-website} explores surveillance issues on the web, and notably the lack of transparency of GTM~\cite{pixeldetracking-gtm-20}.
Finally, a number of books cover the configuration of GTM, attempting to guide website Publishers in their use of the tool~\cite{Webe-2015-Practical-GTM,Chard-2015-GTM-Optimise}.

\noindent
\textbf{Analysis of privacy policies.} 
Multiple works analyzed privacy policies to compare it with actual practices of websites~\cite{libert_automated_2018} and apps~\cite{samarin_lessons_2023}.
These works aimed at discovering discrepancies between the disclosure of the third-parties, types of data collected and cookies, with the actual behavior of websites. 
Libert~\cite{libert_automated_2018} conducted an analysis of the top 25 trackers present on websites, and measured the readability of legal documentation, as well as whether third party trackers are disclosed in privacy policies.
Andow et al.~\cite{andow_actions_2020} focused on Android apps ``leaks'' by automatically comparing privacy-sensitive data flows with privacy policies disclosures and found significant evidence that privacy policies omit as well as incorrectly disclose Data Recipients and data types collected.
Tahaei et al.~\cite{tahaei2020stackoverflow} highlight that developers face difficulties understanding privacy policies and do not always know what to include in these documents. They also state that  
developers may not read privacy policies and consequently may not be aware when they bear legal consequences ~\cite{Tahaei-Developers-2023, tahaei2022advice}.
 
\noindent
\textbf{Summary.} 
Most of the previous works either focus on large-scale measurement of Web tracking, or non-academics describe the functionalities and configurations of GTM and its Tags.
Recent content published by industry experts identify risks for privacy. 
However, no work so far has neither analyzed the behavior of provided Tags, nor the disclosures of data collection in legal documentation of Tags. 
Moreover, there are no previous studies focusing on the ``Google Consent Mode'' specifically introduced by Google to facilitate consent management for GTM Tags. 
These are the gaps that our work aims to fill.

\subsection{Google Tag Manager}
\label{sec:background-gtm}

\subsubsection{Types of Tags in GTM}
\label{sec:background-tag-types}
GTM provides an environment for Website Publishers to choose, configure and include Tags in their websites (see Figure~\ref{fig:gtm-cs-arch}). 
Tags  are in general any HTML content, and are often simply JavaScript libraries that are provided by companies or  developers, whom we call ``\emph{Tag Providers}''. The Tags available in GTM can be put in three categories below. 

\noindent \textbf{81 Natively Supported Tags}~\cite{gtm-official-supported-tags} are shown in the first layer of the GTM interface. These Tags %
are %
for services provided by both Google and third parties. However they do not have a clear description of their Tag Providers. 

\noindent \textbf{702 Community Template Gallery Tags}, that we call ``\emph{Template Gallery Tags}'' are available in a marketplace integrated in GTM~\cite{gtm-template-gallery}.     
Their source code is public and published in a GitHub repository. %
These Tags are provided by an independent developer or company. Such Tags implement a specific service, for example, the independent developer Simo Ahava provided 22 Tags\cite{simo-ahava-templates}, and one of them %
is the Facebook Pixel\cite{ahava-facebook-pixel}. As a result, Tags send requests to third-party servers, and we call receivers of such requests, ``\emph{Data Recipients}''. Therefore, for Template Gallery Tags, Tag Providers are not necessarily the same entities as Data Recipients.

\noindent \textbf{Custom image and Custom HTML Tags} %
are not pre-packaged, but instead 
 are ``placeholders'' that allow the Publisher to integrate any snippets of HTML as Tags. We do not study them further in this paper. 

\subsubsection{Tag installation process}
When a Publisher selects a  Tag, they must first configure it. 
This involves creating an account on the Data Recipient's website so the server can identify the Tag and collect data. 
Next, the Data Recipient validates the Publisher's website before allowing the Tag to be added.
The Data Recipient then provides a unique ID for each Tag and website pair, which the Publisher manually inputs into the Tag via the GTM interface to identify the source of incoming requests.

Once all Tags are selected and configured, GTM generates a unique ``GTM Web Container'', which is a JavaScript library  \gtmjs\ that now contains all the selected Tags (see Figure~\ref{fig:gtm-cs-arch}). 
A small code is then generated to include in Publisher's website - this code will fetch \gtmjs\ and seamlessly include all selected Tags in Publisher's  website at runtime.

When the Data Recipient receives requests from the Tag that the Publisher installed on their website, the Data Recipient provides a specific interface that we call a ``\emph{Dashboard}'' to represent the collected information in a unified form (see an example in  Figure~\ref{fig:dashboard-example} of the Appendix).

\subsection{Google Consent Mode v2}
\label{sec:back-consent-mode}

Google has introduced the ``Consent Mode''~\cite{google-consent-mode-about} --  a mechanism that allows to communicate user's consent status to Google. 
Consent Mode v1, introduced in 2020, contained two consent mode parameters -- \adstorage and \analyticsstorage %
\cite{ga4-consent-types-snapshot2023}. 
This version, now called ``Basic implementation'',   only allowed to block tags from execution when the consent is denied by the user. 
\begin{table}[t]
    \centering
    \caption{Google Consent Mode v2 parameters from Google's official documentation~\cite{ga4-consent-types-snapshot2024}.}  %

    \begin{tabular}{p{2.8cm}p{4.7cm}} %
        \hline
         \textbf{Consent mode \newline parameter}  & \textbf{Description}  \\ 
         \hline
         \adstorage                                       & Enables storage (such as cookies) related to advertising \\
         \cellcolor[gray]{0.9} \analyticsstorage          & \cellcolor[gray]{0.9} Enables storage (such as cookies) related to analytics e.g., visit duration \\         
         \adpersonalisation %
         & 
         Sets consent for personalized advertising \\
         \cellcolor[gray]{0.9} \aduserdata %
         &
         \cellcolor[gray]{0.9} Sets consent for sending user data related to advertising to Google \\   
    \hline
    \end{tabular}
    \label{tab:consent-mode-parameters}
\end{table}

On March 4, 2024, a few days before the EU Digital Markets Act~\cite{DMA} came into force, Google has introduced an updated system --  Consent Mode v2 -- and added two additional consent mode parameters: \aduserdata, and \adpersonalisation~\cite{ga4-consent-types-snapshot2024} (all four up-to-date consent mode parameters %
are shown in Table~\ref{tab:consent-mode-parameters}). %
This version introduced an ``Advanced implementation'' that allows Tags to load and run, and their behavior can be modified when consent is declined. 
According to Google documentation~\cite{google-consent-mode-about}, Tags send ``\emph{cookieless pings}''  that communicate the consent state and other user activity. %
As of now, only Tags provided by Google modify their behavior with respect to the consent given by the user~\cite{google-consent-mode-about}.

\subsection{EU Data Protection Law}
\label{sec:legal-background}

The General Data Protection Regulation (GDPR)~\cite{GDPR} applies to the processing of \emph{personal data}~\cite{EDPB-4-07} and imposes obligations on those actors who determine the data processing purposes and its means. %
The ePrivacy Directive (ePD)~\cite{ePD-09} provides~\emph{supplementary} rules to the GDPR in particular for the use of  tracking technologies. 
Whenever any information %
 is stored and read from the user's device, the ePD~\cite[Art. 5(3)]{ePD-09} requires  
organizations to request user \textit{consent} %
for certain \emph{purposes}, such as advertising~\cite{EDPB-2-13,EDPB-3-13} to process data.
The only way to assess with certainty whether consent is required is to analyze the \emph{purpose} of each tracker on a given website~\cite{EDPB-4-12,Foua-etal-20-IWPE}.

\subsubsection{Personal data} 
\label{sec:personal-data}
Personal data is
\textit{``any information relating to an identified or identifiable natural person ('data subject'). An identifiable natural person is one who can be identified, directly or indirectly}~\cite[Art. 4(11)]{GDPR}. %
In order to determine whether a person is \textit{identifiable}, account should be taken of all the means likely reasonably to be used %
by any actor to identify that person.
Accordingly, if  certain data, alone, is not personal data, it becomes personal data \textit{as regards to someone who reasonably has the  means of enabling that data to be associated with a specific person}~\cite[para. 46]{Scania-case-2023}. %
GDPR Recital 30 asserts that online identifiers provided by their devices, such as IP addresses, can be associated to a person, thus making them identifiable. 
This identification does not require that all the information enabling that person to be identified should be in the hands of a single entity \cite[para. 42, 43]{Breyer2016}.
\textit{Processing of personal data} consists of 
``any operation(s) performed on personal data, such as collecting, sharing, using, making available, accessing, combining''~\cite[Art. 4(2)]{GDPR}. %

\subsubsection{GDPR obligations for Website Publishers} %
\label{sec:gdpr-obligations}

 Publishers can be held responsible and fined if they fail to comply with their obligations under the GDPR~\cite[Art. 28(3)(f), 32-36]{GDPR}.
 When processing personal data, website Publishers must comply with the following GDPR principles, %
 that will guide our legal analysis:
\textit{Lawfulness}: collect personal data only with a valid legal basis (Art. 5(1)(a)); 
\textit{Fairness}: avoid processing data unjustifiably detrimental, discriminatory, unexpected or misleading (Art. 5(1)(a));
\textit{Transparency}: inform users about purposes, recipients, and legal bases when collecting data (Art. 5(1)(a), 13-14); 
\textit{Security}: ensure data protection to prevent unauthorized access (Art. 5(1)(f), 32);
\textit{Data Protection by Default}: process data with the highest privacy settings by default (Art. 25(1)(2);
\textit{Accountability}: demonstrate compliance at all times 
 with data protection principles (Art. 5(2)).

\section{Methodology}
\label{sec:methodology}

In this section, we present our new methodology to analyze the behavior of Tags available in GTM and combine it with the analysis of Tags' legal and technical documentation. 
We then complement this with legal analysis to identify potential violations of the EU laws to answer our research questions from Section~\ref{sec:intro}. 

\begin{figure*}[tb]
    \centering
    \includegraphics[width=0.87\linewidth]{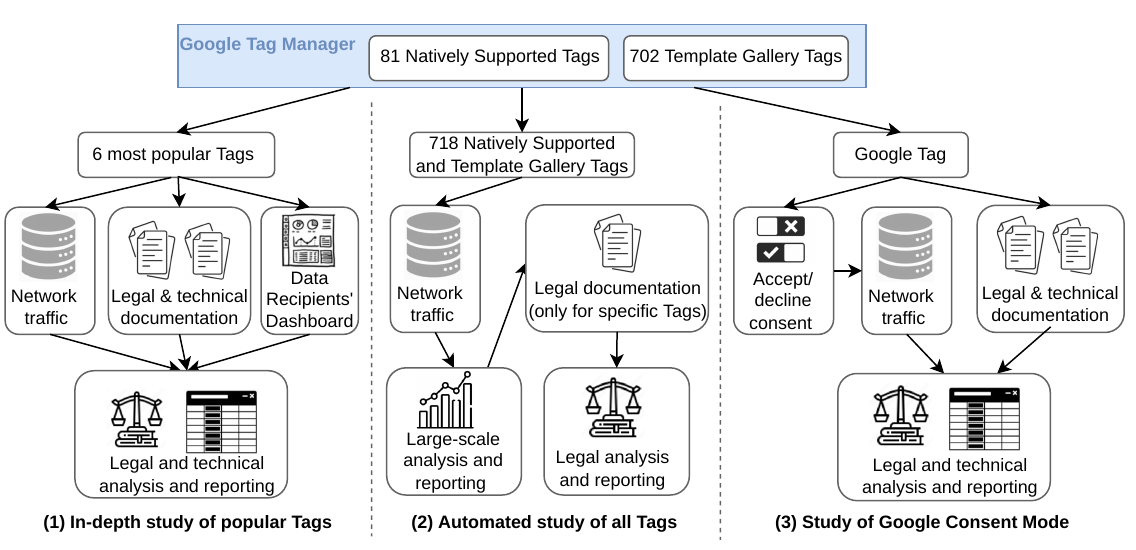}
    \caption{High-level overview of our framework to analyze the behavior of Tags and their legal compliance. 
    }
    \label{fig:pipeline}
\end{figure*}

Our approach is to use the method of \emph{detecting privacy leaks in isolation} for each studied Tag by installing them in a controlled environment instead than in the wild. %
We then analyze the network traffic collected during the Tag's execution
and rely on extensive analysis of the legal and technical documentation of companies that receive data sent by the Tag in the network requests (Section~\ref{sec:meth-legal-tech-docs}).

Figure~\ref{fig:pipeline} gives an overview of three studies we have performed. The first study proposes an in-depth  analysis of popular Tags, %
in the second study  we automatize the process of analysis of Tags' behavior, and the third study evaluates data sent by Google Tag that implements Consent Mode v2. %
Our methods are further detailed in Section~\ref{sec:meth-isolation-method}. 
We ran these studies in March-April 2024 on Google Chrome version~122, from the Flathub repository~\cite{chrome-flatpak}, using the default settings and installed on a GNU/Linux operating system (kernel~6.6.x-lts).
We visited the websites from a computer connected to the Internet through an institutional network in the EU. 

\subsection{Detecting Privacy Leaks in Isolation}
\label{sec:meth-isolation-method}

Prior research showed that third-party libraries are rarely interacting with only one tracking domain~\cite{Foua-etal-20-PoPETs,Engl-etal-16-CCS}. 
Instead, such libraries often include other content, %
thus initiating %
requests to multiple domains, and often enabling tracking the user. 
Therefore, analyzing Tags' behavior on the websites, where they are installed, may be very complex due to the presence of multiple Tags and communications with tens of servers on the same page. 

To overcome this problem, we \emph{detect privacy leaks in isolation}. This method is inspired by the previous research on analyzing the behavior of Consent Management Platforms (CMPs) in a fully controlled environment~\cite{toth_dark_2022,Sant-etal-21-APF} by installing them one-by-one on an empty website and evaluating their behaviors in isolation. 
Our method is also related to  the ``unitary asset testing methodology''%
in software development, when each asset (in our case, each Tag) is configured and run one at a time, in a fully controlled execution environment to collect all artifacts generated during its execution. 

\subsubsection{Creating controlled environment}
To analyze the behavior of each Tag in isolation, we assume the role of a website Publisher, whose goal is to configure GTM and select Tags the Publisher wants to include in their website.%
To simulate such Publisher who sets up the GTM infrastructure, we bought a TLD+1 domain that we call \texttt{example.com} in the rest of the paper.
We host \texttt{example.com} on a virtual machine we rented on a public cloud provider and located in France. 
We created a new Google account to create GTM Web containers to install on our own websites (Web container is needed to install Tags in GTM, see Figure~\ref{fig:gtm-cs-arch}). 

For an in-depth study, we created  a new  Web container for each Tag to be tested and installed it on a new corresponding website \texttt{<tagID>.example.com}, where \texttt{tagID} corresponds to a unique identifier of a Tag in GTM. 
For our automated study, we use a single website \texttt{example.com} and Web container where our automated script installs each Tag in isolation one after the other. %
Our in-depth and automated analyses did not use consent management solutions.
We kept the default configuration for our \texttt{Nginx} web server, which does not set security headers such as \texttt{content-security-policy} or \texttt{Access-Control-Allow-Origin}. 
This allowed the website (and its tags) to send HTTP requests to any third-party domain. 
The source code of our website is available in the supplementary materials\cite{supplementary-materials}.
We used the ``profiles'' functionality of the browser to make every visit in a fresh environment, devoid from cookies, local storage and other technologies than maintain a state. 

\subsubsection{Capturing network traffic}
\label{sec:capturing-traffic}
To study the data collected and sent by Tags in both in-depth and automatic studies, we used the developer tools integrated in the Google Chrome browser during a new visit and used the export function that creates a HAR~\cite{har} report containing HTTP(S) requests and responses as well as cookies stored on our empty website.  
In order to have a reference point, we first install an empty Web container on our website and capture all the network traffic in a form of a HAR report. We call it \emph{``baseline traffic''}. It contains  HTTP(S) requests to our website (\texttt{<tagid>.example.com} for in-depth study and \texttt{example.com} for automatic study) and a request to \texttt{googletagmanager.com} loading the GTM Web container. 
When analyzing traffic generated by each Tag, %
we filter out the baseline traffic to obtain only traffic generated by the Tag.

\subsubsection{Official and Unofficial Tags} 
\label{sec:official-tags}

\begin{figure}
    \centering
    \includegraphics[width=0.85\linewidth]{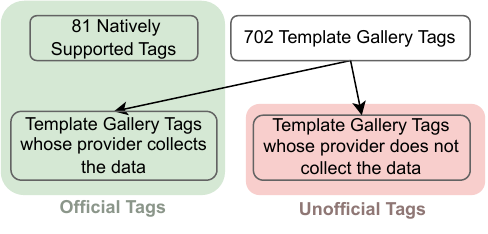}
    \caption{Official and Unofficial Tags in GTM.}
    \label{fig:official-tags}
\end{figure}

Sometimes, the company or developer that provides a Tag -- the \emph{Tag Provider} -- is also the receiver of requests initiated by the Tag -- the \emph{Data Recipient} (Section~\ref{sec:background-gtm}). %
However, %
some Tags are also provided by independent developers who implement services for different Data Recipients. 
Consequently, in our study, we propose to separate Tags into two groups that we call ``Official''  and ``Unofficial'' (shown  in Figure~\ref{fig:official-tags}).

We consider a Tag to be Official when its Tag Provider is also a Data Recipient (but since some Tags send data to many different servers, there could be several Data Recipients); otherwise, we consider the Tag to be Unofficial.

Natively Supported Tags are provided by the companies that implement their own service, and we note that the names of these Tags contain the name of the company.
We therefore assume that these Tags are provided by the companies that collect the data, meaning that Tag Providers are also Data Recipients.
We consider all such Tags to be Official. 

Template Gallery Tags instead require further detailed analysis.
To determine who is the Tag Provider, we visit its GitHub repository associated to the Template Gallery Tag and leverage GitHub's organization verification feature~\cite{github-organization-verification} that certifies that a certain company is the owner of the repository. 
We then consider this company to be the Tag Provider. 
If there is no GitHub certification for the repository, we consider such Tag to be Unofficial. 
However, it is not possible to %
know who are the Data Recipients by analyzing the publicly available information about Template Gallery Tags. 
We therefore analyze the traffic initiated by each Template Gallery Tags to identify whether the Tag sends requests to its Tag Provider.
If yes, we conclude that the Tag Provider is also a Data Recipient, and therefore such Tag is considered Official.

\subsection{Legal and technical documentation}
\label{sec:meth-legal-tech-docs}

Previous research on data collection disclosures focused solely on analyzing the privacy policies of websites and apps~\cite{andow_policylint_2019,andow_actions_2020}.
Other works \cite{Dege-etal-19-NDSS, libert_automated_2018} %
included 
cookie policies and terms of use.
While conducting a preliminary analysis of privacy policies of Data Recipients, we found that disclosures of data types spreads across %
even more documents, differing among the Data Recipients, such as technical documentation. 
Therefore, we took a different approach than prior work, and included %
these documents, rather than limiting ourselves to legal documentation. 

For the study of the documentation, we %
chose to analyze only Official Tags, when the Tag Provider  is also Data Recipient.  %
We  searched for  both  legal and technical documentation of Data Recipient that could be available to the website Publisher who is installing the respective Tag. 
We collected the sources of information on the Data Recipient's website and also during the sign up process when creating an account on its website. %

The set of \emph{legal documentation} was defined together with the legal scholar, and it contains  \emph{policies} (e.g.,  ``Privacy/Advertisement/Cookie Disclosure Policy''),
\emph{terms} (e.g., ``Terms of service'', ``Terms of use'', ``Product terms'', ``Data Processing Terms''), 
\emph{agreements} (e.g., ``Data sharing agreement'', ``Service Agreement'', `` Data Processing Agreement'') and 
linked \emph{addendum} (e.g., ``Data Protection Addendum''). For the Google Tag, we have studied the legal documentation of the Google Analytics  that has recently been included in the Google Tag~\cite{google-analytics-becomes-google-tag}.

By \emph{technical documentation}, we consider all other documents  provided by the Data Recipient, such as technical guides, FAQs and information for developers.

We found that legal documents are often broad, covering multiple products and the Data Recipient's website. Our legal expert could not identify the relevant sections alone, so the reading was done collaboratively by the legal expert and a computer scientist. Time spent on each Tag's documentation was limited to 2 hours.

After collecting all documentation for the Tags in each study, neither the legal scholar nor the computer scientists could identify which parts \emph{specifically referred to a given Tag}.
This occurred for 
4 out of 6 studied Tags in the in-depth study,
10 out of 11 Tags selected in the automated study, 
and for Google Tag in the consent mode study. 
We believe the lack of clarity arises because the legal documentation is company-wide, not product-specific. Additionally, terms like "visitors" are ambiguous, as they could refer to either Publishers visiting the Data Recipient's website or users visiting the Publisher's website where the Tag is installed.
Related work on privacy policies\cite{pp-across-ages} reports this same \textit{lack of specificity} of finding %
data practice descriptions %
specific to a given website or service; they are often quite generic as they apply to a range of products of that organization or are umbrella policies.
\subsubsection{Limitations}

We have analyzed the collected documentation searching for specific disclosures, while acknowledging that the lack of specificity may impact the results presented in this paper.
Our analysis of documents is also limited by one IT and one legal expert which might have introduced bias in their interpretation.

\subsection{In-depth study of popular Tags}
\label{sec:meth-in-depth-study}

In our first experiment, we perform an in-depth analysis of popular Tags by installing them one-per-website, each on a dedicated domain \texttt{<tagID>.example.com}. Our goal is to study each Tag's behavior and detect privacy leaks by performing actions a typical website Publisher would perform. 
In this study, we manually interacted with the website, prompting the tags to collect data.
We performed the following actions: moving the mouse, downloading a file, scrolling, and using the search bar on our website.
We contrast and compare three sources of information available to the Publisher:
\begin{itemize}
    \item {\em Network traffic:} users’ data transmitted by the Tag that we capture in the network traffic initiated by the Tag (Section~\ref{sec:meth-isolation-method});
    \item {\em Legal and technical documentation:} information about data collected by the Data Recipient disclosed in their official privacy policies and other technical documents (Section~\ref{sec:meth-legal-tech-docs});
    \item {\em Dashboard:} information about data received by the Data Recipient shown to the Publisher in the Dashboard (Section~\ref{sec:background-gtm}).
\end{itemize}

\subsubsection{Selecting and installing popular Tags}
\label{sec:meth-selecting-tags}
In order to be able to compare the behavior of a Tag to its own documentation, we decided to study only Official Tags, where the company that provides the Tag is the same as the company that receives the data from the Tag, that is when Tag Providers are the same as Data Recipients (Section~\ref{sec:official-tags}).
This solution allows us to study the legal and technical documentation of the Tag Provider, which would not be possible if the Tag Provider was a different entity than the company collecting the data. 

In order to select the most popular Tags, we first identify the most popular third-party domains based on prior work, and then match these domains, that represent Data Recipients, to the Official Tags. 
We use the top~50 most prevalent third-party domains collected from Alexa top 10K in 2020~\cite{Foua-etal-22-PETs}. 
For each domain found, we use Tracker Radar~\cite{tracker-radar} to identify the company who owns the domain.
Out of the top 50 third-party domains in our list, we identified 30 distinct companies that own these domains. 
We then pick one Official Tag proposed by the company we identified. %
For example, for the most popular third-party domain \texttt{google-analytics.com}, we identified its owner Google and selected the ``Google Tag'' since it is offered by Google and comprises the Google Analytics functionality in this Tag~\cite{google-analytics-becomes-google-tag}. 
Whenever we do not find any Tag proposed by the company behind a third-party domain (e.g.,, there is no Tag that contains ``Microsoft'' in a Tag's name), we search for the Tag whose name contains the domain name (we have searched for \texttt{linkedin.com} and found the Tag named ``LinkedIn Insight'').

We have identified that out of 30 companies that own 50 top third-party domains, only  8 companies propose Tags in GTM\footnote{%
The top~50 most popular third-party domains with the corresponding Tags we identified are available in the Appendix, Figure~\ref{fig:top-domains-tags-mapping} and in the supplemental materials~\cite{supplementary-materials}.%
}. 
We have installed each of these 8 Tags in the corresponding website \texttt{<tagID>.example.com} and tried to create an account on the website of the Data Recipient, so that its server recognizes the specific requests it will receive from this Tag and properly collect received data. 
However, Data Recipients often check and validate the Publisher’s website before allowing the Tag to be added (Section~\ref{sec:background-gtm}). 
In our experiment, for 2 Tags -- ``Criteo OneTag - Official'' and ``LiveRamp'' -- our  account was not validated, and we therefore proceeded with the remaining 6 Tags, shown in Table~\ref{tab:selected-tags-popularity}. 

\begin{table}[t]
    \centering
            \caption{The 6 Tags selected for in-depth study. A ``*'' indicates Tags selected from the Template Gallery, the rest of the Tags are Natively Supported by Google. }

        \begin{tabular}{P{1.8cm}P{1.7cm}P{3.4cm}}
            \hline %
             \textbf{Tag selected}                & \textbf{Domain owner from~\cite{tracker-radar}}  & \textbf{Domain and ranking according to~\cite{Foua-etal-22-PETs}} \\ \hline %
             Google Tag                           & Google LLC                                               & \texttt{google-analytics.com}  (1)                                \\ \hline %
             Twitter Base Pixel *                 & Twitter, Inc.                                            & \texttt{twitter.com}           (15)                               \\ \hline %
             comScore Unified Digital Measurement & comScore, Inc                                            & \texttt{scorecardresearch.com} (18)                                 \\ \hline %
             Quantcast Advertise                  & Quantcast Corporation                                    & \texttt{quantserve.com}        (21)                                 \\ \hline %
             Hotjar Tracking Code                 & Hotjar Ltd                                               & \texttt{hotjar.com}            (27)                                 \\ \hline %
             LinkedIn Insight                     & Microsoft Corporation                                    & \texttt{linkedin.com}          (46)                                 \\ \hline %
        \end{tabular}
        \label{tab:selected-tags-popularity}
\end{table}

\subsubsection{Dashboard of the Data Recipient}
While installing each Tag, we create an account on its Data Recipient's website to then add the Tag in its corresponding Web container. 
We follow instructions to configure it, and  ``link'' the Tag to our account.
Once the Tag is installed and configured on the corresponding \texttt{<tagID>.example.com} website, we visit this website and then log on to the Data Recipient's website, to see  a specific interface that we call ``\emph{Dashboard}'', which  represents the collected information in a unified form (see an example in Figure~\ref{fig:dashboard-example} of the Appendix).
Finally, we identify in the Dashboard the data that the Data Recipient has received when we visited the Tag's website \texttt{<tagID>.example.com}.
The proof that the Data Recipient actually has registered the data types from the incoming network traffic is the demonstration of such data types in the dashboard. 

\subsubsection{Network traffic analysis}
\label{sec:in-depth-traffic}
There is no specific documentation on what data is sent by Tags in the network traffic, and where exactly in the traffic it is located. 
We therefore first build a list of all possible data types by analyzing the  legal and technical documentation as well as the  dashboard for the 6  Tags. 
The resulting list can be found in the first column of Table~\ref{tab:in-depth-data}.
For each data type in the list (e.g., ``screen resolution''), we detect when the value of the data type (i.e., our screen resolution) was shown in the dashboard of a given Tag. We then search for this value in the traffic initiated by the given Tag.
Specifically, we  search through 
URL parameters of the HTTP(S) requests,
body of the HTTP POST requests and 
content of the WebSocket messages that are sent. 
For example, in Google Tag's dashboard, we identified that our screen resolution was collected (shown as ``2560x1440''). We  searched for the two values, ``2560'' and ``1440'' in the HAR file we collected on our website \texttt{googletag.example.com} where Google Tag is installed. We found a ``2560x1440'' value in one of the URL parameters in a request to \texttt{google-analytics.com}.
When analyzing \textit{Hotjar Tracking Code} Tag, we found in its dashboard that the screen resolution was also collected. 
However, when analyzing network traffic initiated by the \textit{Hotjar Tracking Code} Tag installed on \texttt{hotjar.example.com}, we found that it sends the window width and height in two separate parameters in the JSON format sent via a WebSocket. 

With our method, for each Tag, we determine a list of data types \emph{actively collected by the Tag} and sent in its network traffic. 
The active collection of data types by Tags demonstrates that the data is intentionally sent by the Tag's script and therefore collected by Data Recipients.

Because of the observed diversity of formats and location in sending the same data type across Tags (e.g., the “screen resolution”), it is hard to unify the analysis and detection of data types uniformly among all Tags and we proceed one Tag at a time.
Nevertheless, we noted that the referral URL (URL of the previously visited page), the page title and the page URL are usually sent in clear. Therefore we have opted for also detecting these three data types across all Tags uniformly by searching for these three values in clear in the outgoing requests. 
Whenever we find these data types in the traffic, we add them to the list of data types actively collected by the Tag. 

Additionally, we record all the data types that are \emph{passively sent by each Tag}, that is, IP address, TLS session ID, and HTTP headers that are sent with  HTTP requests  (user-agent header: browser name/version, device type/name, accept-language header: browser language, referrer header: page URL). 
Since this data is sent passively, we further consider that the Tag has indeed collected this data only if such data type appears in the dashboard, or in the legal or technical documentation, indicating that the this data type was indeed collected.

\subsubsection{Limitations}
\label{sec:meth-in-depth-limitations}
We do not analyze prevalence of tags in the wild and therefore we do not provide further statistical results about the data that the Tags collect in the wild. 
However, our in-depth study does demonstrate the types of data.
We limited the number of Tags because it is impossible to properly install and configure them automatically (Section~\ref{sec:meth-selecting-tags}). 
Indeed, across the Tags we installed, the configuration processes vary a lot, dashboards present information in very  different forms, and legal documentations are organized differently. 
Among the 6 analyzed Tags, ``\textit{comScore Unified Digital Measurement}'' and ``\textit{LinkedIn Insight}'' did not let us see the dashboard, either because our website did not pass a manual review (comScore), or because the Tag did not collect data from a sufficient number of visitors (LinkedIn).
Tags which collaborate to collect data (e.g., one gathering the information and the other exfiltrating it) are not studied as our methodology consists in analyzing tags individually.
While we studied technical documentation and included elements Tags rely upon to collect data, our website did not include all the features found on public websites (e.g., login form, or product basket) and Tags could collect more data than we observed.

\subsection{Automated study of all available Tags}
\label{sec:meth-automated}

In the second study, we aim at analyzing all the Tags available in GTM. 
However, a proper installation and configuration of a Tag requires to set up an account on the Data Recipient's website and ensure that further setup process is complete. Our analysis of 6 popular Tags in the in-depth study 
showed that such processes differ significantly across Tags and therefore are hard to automatize. 

\subsubsection{Automated lightweight installation and data collection}
\label{sec:lightweight-installation}
To scale Tag analysis, we used a simplified installation process. 
Since automating account creation and Tag configuration on the Data Recipient's website is complex, we developed a script to automatically install Tags and fill in required fields with invalid but plausible information.

We used the GTM API to automatically install the 81 natively supported tags. However, this API does not support the installation of template gallery tags. 
We therefore automatically imported the 702 Template Gallery Tags from their source code, available on the associated Github repository, that is indicated in the description of each such Tag in the GTM interface.  %
Interestingly, we found that Github source code is sometimes more recent than the code available in GTM (i.e., the Tag Provider pushed new code to Github but GTM did not update yet the Tag in the template gallery). Also, we found Tags that contain errors -- the Tag Provider pushed invalid code to Github and GTM could not update the Template Tag -- and we discard such Tags from further analysis.

When adding a Tag to a GTM Web Container, its configuration fields are often validated by GTM.
If any field is invalid, the GTM Web Container throws an error and the Tag is not saved.
Our script initially attempts to leave all fields blank when adding the Tag.
In case of errors, the script generates a string that meets the requirements of the specified error and retries to install the Tag.
Examples of such errors are ``value must not be empty'', ``value must be a positive integer'', ``value does not match the regular expression'', ``length must be between x and y characters''.
The code generating the configuration strings is available in the artifacts.

As a result, our script -- available in the supplemental materials~\cite{supplementary-materials} -- makes the following process for each Tag. 
First, it automatically installs a Tag on our empty website \texttt{example.com} by using GTM's API~\cite{gtm-api}. %
Second, the script visits our website \texttt{example.com} with Puppeteer~\cite{puppeteer} version 22.6.4 to collect the HTTP requests and responses in a form of HAR~\cite{har} as well as cookies installed by the studied Tag. 
Finally, it closes the browser and uninstalls the Tag.
Out of 81 Natively Supported Tags and 702 Template Gallery Tags, we managed to automatically install and analyze \tagstotal\ Tags (that is, 74 Natively Supported  and 644 Template Gallery Tags). 

\subsubsection{Network traffic analysis} 
\label{sec:meth-automated-traffic-analysis}
We use the \emph{baseline traffic} we have collected by installing an empty Web container on our website \texttt{example.com} (Section~\ref{sec:capturing-traffic}). 
For each Tag, we collect the traffic in the form of HAR file~\cite{har} and cookies set during the automated visit to \texttt{example.com} where the Tag is installed. 
We then remove the baseline traffic in the HTTP %
traces to obtain only traffic generated by the Tag. 
Since no cookies are present in the baseline traffic, %
we do not apply such filtering to cookies.

We extract from the HAR file all TLD+1 contacted  by the studied Tag and identify the companies that own these domains using the Tracker radar~\cite{tracker-radar}. Whenever we do not find the domain, we also use the Disconnect list~\cite{disconnect-me}. %
When we do not find a domain in neither list %
we consider each unidentified TLD+1 to be a different company. 
As a result, for each of the \tagstotal\ Tags, we obtain the list of domains/companies contacted by the Tag together with the cookies that these domains install.

\subsubsection{Legal documentation}
\label{sec:meth-automated-legal}

For Tags that contact multiple companies, we further need to check the \emph{legal documentation} of the Tag Providers, to see whether they properly disclose the potential sharing of users' data.
In this work we only analyze the legal documentation of official Tags, for which the Tag Provider is also Data Recipient (Section~\ref{sec:official-tags}).
So we search for the legal documentation on the website of the Data Recipient, and check whether: it 
clearly mentions the Tag;   
it declares all the companies that this Tag is contacting according to our network traffic analysis; 
and whether third-party cookies created by the Tag are described in the legal documentation.

\subsubsection{Limitations}
\label{sec:automated-study-limitations}

Our crawler did not interact with our website.
As a result, %
we do not observe the behavior of Tags that are triggered by user interaction.  
Since we configured Tags with placeholders instead of linking them to an account on Data Recipient's websites, 
we may have missed specific Tags behavior that would happen only when Tags are configured with the proper account.

Some Tags could not be installed automatically because our script could not configure the Tag with plausible information.
Here also, our testing method misses all tags that are collaborating (e.g., one gathering the information and the other exfiltrating it) as every Tag is only tested in isolation.
In this experiment, observations should be considered as a lower bound of data collected by Tags.

We did not analyze the technical documentation for Tags that communicate with multiple companies, as it is typically available only after account creation, which we did not do in this study.

\subsection{Comparative study of Tag behavior between our in-depth and automated studies}
\label{sec:meth-comparison}
To better understand the implications of our methods, we compared the behavior of Tags common to the two studies using our two methods: in-depth and automated.
To achieve this, we compared the network traffic of 6 tags using these two methods.
In the automated method, only one page is visited, while the in-depth method involves two pages due to interaction. For a fair comparison, we only consider traffic from the first page in the in-depth method and compare it with the automated method.
We first compare the number of successful requests by type: script downloads, GET requests (except scrips), POST requests and WebSockets opened.
Then, we search for the data types identified in the in-depth setting, in the network traffic collected within the automated setting.

\subsection{Study of Google Consent Mode v2}
\label{sec:meth-consent-mode}

The Consent Mode v2 introduced by Google allows to communicate the consent decisions from the consent banner provided by the Consent Management Platform (CMP) to the Tags (Section~\ref{sec:back-consent-mode}). Consent Mode introduced four ``consent parameters'' (see Table~\ref{tab:consent-mode-parameters}) that can have two possible values,  ``granted'' or ``denied''~\cite{manage-consent-settings-snapshot-june2024}. 

The five Tags provided by Google (Section~\ref{sec:back-consent-mode})
contain checks for the values of consent mode parameters in their code, and modify their behavior based on the values of these parameters, thus providing the ``Advanced Implementation'' of  Consent Mode v2. To analyze the consent mode, we have selected the ``Google Tag'' since it now contains the popular Google Analytics functionality~\cite{google-analytics-becomes-google-tag}. 

To analyze how the Google Tag changes its behavior depending on consent provided by the user, we have set up two pages %
\texttt{consent.html} and \texttt{noconsent.html} on \texttt{consent-mode.example.com}.
These pages contain a text, a search bar and a link to download a file.
To each page, we add the code provided %
to CMPs to set the default  value of consent variables~\cite{gtm-set-default-consent} to test two settings: 

\begin{itemize}
    \item \emph{refuse all}, where all the consent mode parameters are set to ``denied'' on \texttt{noconsent.html}, and 
    \item \emph{accept all}, where all the consent mode parameters are set to ``granted'' on \texttt{consent.html}. 
\end{itemize}

On each of the two pages, we add the same Web container with only Google Tag installed. 
We then visit each page with the clean browser profile and perform out three actions, simulating user behavior: we download a file, scroll to the bottom of the page, and use the search bar with the therm "my search". For each webpage, we collect the HAR file and cookies.

\subsubsection{Network traffic analysis} 

Our in-depth analysis of Google Tag revealed that this Tag sends user information only in the URL parameters of HTTP(S) requests. 
During our visits on the two pages, after removing the baseline traffic, 6 HTTP requests were made by  Google Tag to \texttt{google-analytics.com}. 
We extracted the URL parameters for each of the 6 requests and compared the types of data sent when consenting and when refusing.
Whenever %
the Tag sets a first-party cookie via its script, we search for this cookie value in clear in the URL parameters of all 6 requests to detect when cookies are sent to Google.

\subsubsection{Legal and technical documentation}
Since Google Tag was already present in our in-depth study (Section~\ref{sec:meth-in-depth-study}), here we compare and contrast our findings from the network traffic for \emph{refuse all} and \emph{accept all} settings to both legal and technical documentation and also additionally study documentation about Consent Mode~\cite{consent-mode-reference,about-consent-mode-modeling-snapshot-june2024}.

\subsection{Proactive prevention of harm}

This research does not involve any human nor the processing of personal data.
Our testing websites are not mentioned in any public sources, and thus are very unlikely to be accessed by the general public. 
We visit these websites with a clean browsing profile, thus not revealing any personal information of the experimenter.

We study services provided by several companies by following these services' legal and technical documentation, analyzing the network traffic on our website and dashboards provided by the services. 
Our research did not involve any written or live discussion with employees %
of these companies, we only used documentation and automated web services.
In this research, we did not find any security vulnerability. However we identified \emph{potential legal violations}\footnote{We underline the importance of using the word ``potential'' since only court decisions can ultimately conclude about its presence.} under EU Data Protection law. 
If we did not seize our IRB for all the reasons mentioned above, we keep the president of our institutional IRB aware of the present work and of discussions with these companies.

After submitting this article, we reached out to 
(1) all 6 companies identified in our in-depth study (see Table~\ref{tab:in-depth-data}) with potential disclosure issues; 
(2) 9 companies whose tags transmit data to multiple third parties without disclosure (see Table~\ref{tab:automatic-11-official-tags}), and 
(3) Google whose Tag collects data without consent (Section~\ref{sec:consent-mode-cookieless-leaks})\footnote{A copy of the messages we sent can be found in the Appendix%
.}.

On November 29, 2024, we have sent 12 emails and 3 paper mail letters to companies when we did not find an email address.
As of March 4, 2025, out of 15 companies we have received 5 responses: 3 reviewed their products, while 2 provided more documentation and invited us to review our findings.
However, reviewing their documentation did not affect the findings reported in the paper.

\section{Findings}
We present our main findings
for the three studies %
(see Figure~\ref{fig:pipeline})
and their legal consequences.

\subsection{In-depth study of six Tags} 
\label{find:in-depth}

For the six Tags we have successfully installed and configured (see Table~\ref{tab:selected-tags-popularity}), we have further analyzed the network traffic initiated by these Tags and compared the data extracted from the traffic with the dashboard, legal and technical documentation of the corresponding Data Recipients (Section~\ref{sec:meth-in-depth-study}).
In this process, we have answered the research questions \textbf{RQ1} and \textbf{RQ2} and identified two major issues: collection of personal data and insufficient disclosures about such data collection.

Within network traffic, we searched for the data type values found in the documentation or dashboard. %
Almost all such values were found in clear text and we therefore did not search for encoded/hashed values. Regarding the only two values we did not find\footnote{See empty cells in ``network traffic'', where the corresponding cells in documentation or dashboard columns are filled in Table~\ref{tab:in-depth-data}.}: first, a referral URL for \textit{Linkedin Insight} was not found in clear text, while non-understandable values changed with each experiment repetition, and such data cannot be the referral URL; second, out of 12 data types reported by \textit{Quantcast Advertise}, only screen resolution collection was not found, that may have been encrypted. However these two cases did not justify a further modification to our methodology.

\begin{table*}[ht]
    \centering
    \caption{ \textbf{Types of data collected according to the network traffic, dashboard, technical and legal documentation}. %
    Numbers in the table %
    correspond to the disclosure issue discussed in Section~\ref{sec:in-depth-disclosure-issues}. 
    For network traffic, a filled cell (with either ``x'' or a number) indicates that the data type has been \emph{actively collected}. %
    The %
    star ``*''  indicates that the %
    data is sent as a part of the HTTP(S) 
    communication, and therefore %
    has been \emph{passively sent}. 
    The ``x'' sign in the dashboard or documentation means that the data type is present. 
    A filled cell in gray areas indicates that the documentation declares this broad category of data being sent. %
    ``comScore UDM'' stands for comScore Unified Digital Measurement.
     }
    \includegraphics[width=0.79\linewidth]{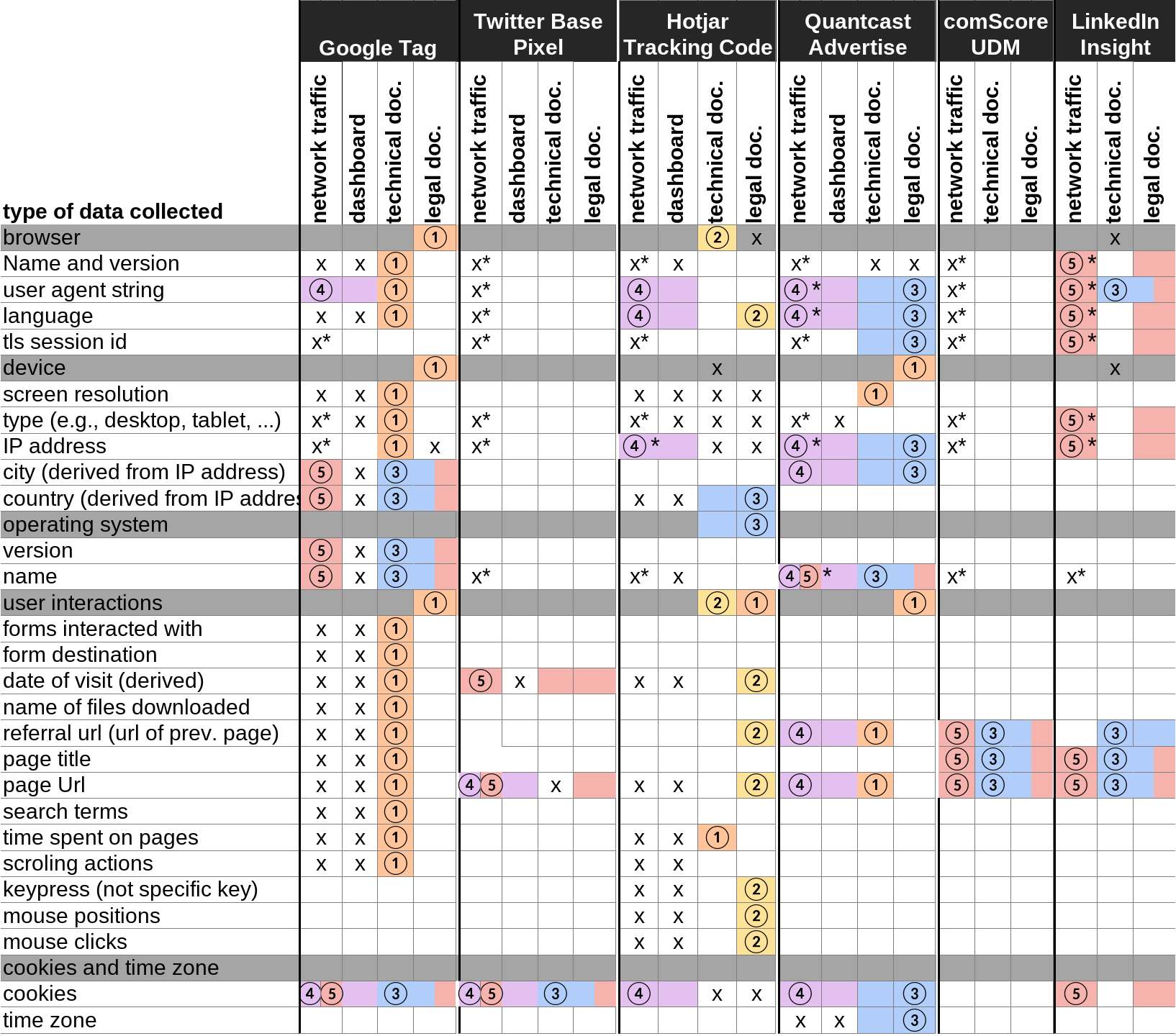}
    \label{tab:in-depth-data}
\end{table*}

\subsubsection{Data collected by Tags can be personal data}
\label{sec:in-depth-data-personal}

In this section, we answer %
\textbf{RQ1} on what types of data the Tags collect and whether such data constitutes personal data. 
The types of data sent by the studied 6 Tags to the Data Recipients contain a variety of information, such as  browser and device information, operating system properties, capturing user interaction, as well as cookies and time zones. %
These data types are presented in Table~\ref{tab:in-depth-data}, alongside with the presence of this data in the network traffic, dashboard, technical and legal documentation. 
We argue that these data consists of \textit{personal identifiable data}. 

Upon a user's visit to a website embedding a Tag, this Tag sends HTTP(S) requests to the Data Recipient's server, and  inevitably, the server is able to access the IP address of the end user.
An IP address would be personal data if, in combination with other data, it relates to an identified or identifiable person\footnote{%
Not every IP address %
is by default personal data and each case should be studied %
by regulatory authorities. The EDPB acknowledges that an IP address could be originating from a user's router %
or a CGNAT grouping a number of subscribers under the same public IP address\cite{EDPB-ePD-2024}.
}. 
In order to determine whether a person is identifiable, account should be taken of all the \textit{means} that can reasonably be used by any actor to identify that person (Recital 26 GDPR). 
This means that any actor having the means to identify a user, renders such a user identifiable~\cite{Breyer2016,PD-29working-party}. %
Data Recipients have the means to collect the end-user IP addresses and to combine it with all information sent together with the IP address by each tag and that relates to an identifiable person, such as browser, device, operating system, user interaction, user information and identification. This combination of indirectly identifiable data is personal data.

\subsubsection{Disclosures of collected data from Data Recipients}
\label{sec:in-depth-disclosure-issues}
We found that Data Recipients declare the data they collect in their technical documentation and/or in the legal documentation, and sometimes demonstrate in the dashboard. Table~\ref{tab:in-depth-data} summerizes our findings, organized by five types of mismatches between the data sent in network traffic, dashboard and documentation. 
Our findings demonstrate that Publishers who need to know what data types are collected, must read and understand \textit{all} the documentation before installing the Tag on their website, to know what data is shared with the Data Recipients. In the rest of this section, we answer the research question \textbf{RQ2} whether Tags disclose what data they collect data in a compliant way. 
As mentioned in Section~\ref{sec:meth-legal-tech-docs} we analyzed, both legal and technical documentation to understand where and how Data Recipients disclose what data is collected by their Tags. 
We present our findings, labeled by numbers, such as {\small{\TgenLspec}}, that can be found in Table~\ref{tab:in-depth-data}, each highlighted in a different color.

\noindent \textbf{\LgenTspec\ Legal documentation is general but technical documentation mentions specific data types.}
Unspecified and general data is disclosed in the legal documentation while technical documentation is specific.  
Google Tag seems to use this as a general strategy, since most of data it sends is mentioned in a general manner in the legal documentation, while technical documents list specific data types. 
\textit{Quantcast Advertise} Tag does not seem to employ a general strategy, it discloses device and user interactions categories in the legal documentation (such as ``information about your interactions with the content''), while its technical documentation lists specific data types, such as screen resolution, referral URL and page URL).

\noindent \textbf{\TgenLspec\ Technical documentation is general but legal documentation mentions specific data types.}
We also identified a disclosure problem opposite to issue 
\LgenTspec , that is specific to \textit{Hotjar Tracking Code} Tag. %
It actively collects the language, referral and page URL, keypresses and mouse positions and clicks, and derives the date of the visit. All these specific data types are  mentioned in the legal documentation, while  the technical documentation only mentions at a general level that browser information is collected.

\noindent \textbf{\EitherLorT\ 
Data types are declared in either  legal or technical documentations, but not in both.} 
We observed that data collected is %
disclosed only in legal or technical documentation, affecting all six Tags studied.
Collected data is disclosed in the technical documentation but absent in the legal one 
for Google Tag regarding derived city and country, and version and name of the operating system. 
 Conversely, Hotjar declares  country and OS in legal documentation but not in the technical one.

\noindent \textbf{\NotinDash\ Some Data Recipients do not show collected data in the dashboard.}
By analyzing the dashboards of four Tags -- where the dashboards were accessible\footnote{\textit{comScore Unified Digital Measurement} and \textit{LinkedIn Insight} Tags did not allow us to see the dashboard, see Section~\ref{sec:meth-in-depth-limitations}.}-- we found that, for all Tags, some data sent in the traffic is not shown to the Publisher on their respective dashboard. 
These four Tags set cookies. For example, Google Tag and \textit{Hotjar Tracking Code} Tags set first-party cookies, and \textit{Twitter Base Pixel} and \textit{Quantcast Advertise} Tags set third-party cookies -- however, no cookie-related information is  shown in the Data Recipient's dashboards.

\noindent \textbf{\TrafnotinL\ 
Data is sent in network traffic but not declared in legal documentation.} 
Five out of the six studied Tags send specific data types that are not disclosed in the legal documentation. 
Google Tag passively sends the IP address of the user, as stated in the legal documentation.  
Additionally, Google Tag seems to derive the city and the country from the user's IP address, and while these data types are shown in the dashboard and mentioned in the technical documentation, such derived information is not mentioned in the legal documentation. 
From the network requests of the \textit{Twitter Base Pixel} Tag, Twitter derives the date of visit and shows it in the dashboard, but does not declare it in any documentation. This Tag also does not disclose the collection of third-party cookies.
\textit{comScore Unified Digital Measurement} and \textit{LinkedIn Insight} Tags send the page title and website URL and mention it in technical, but not in legal documentation.

\subsubsection{Consequences for Publishers}

Publishers need to read \emph{both legal and technical documentation} to understand data collection practices associated with the Tags they include, however none of these documentations is complete 
\LgenTspec\:, \TgenLspec\:, \EitherLorT.
Moreover, previous works assert that legal documentation is difficult for non-lawyers to understand, and that  developers have insufficient knowledge of privacy policies~\cite{tahaei2020stackoverflow, Tahaei-Developers-2023, tahaei2022advice}. %
An alternative solution could be to use the \emph{dashboards}, but we question their value  and purpose %
\textbf{\NotinDash}, since 
Publishers cannot depend on these as a transparency resource because certain data is collected, %
but never shown on the dashboard.
Alternatively, Publishers could take on the burden of conducting their own data traffic analysis.
However, such traffic analysis shows that the disclosure of data types in the legal documentation is not enough for Publishers to understand what data is, in effect, collected by the Tags they use \TrafnotinL. 
As a result, Publishers may struggle to determine the types of collected data, and whether data is being derived or passively sent. 
This fact denotes omissions between the actual data flow and the legal documentation (also known in the literature as \textit{flow-to-policy omissions}~\cite{andow-privacy-policy-2020}). 
Additionally, performing  traffic analysis presents another challenge, as it requires technical expertise of network principles and specific tools, that is at odds with the overall approach of GTM aimed at an easy integration of third-party Tags.

\subsubsection{Compliance Consequences} 
Data Recipients shift the burden onto Publishers to understand their own data collection practices through the tags they build. In particular, Publishers need to understand how legal and technical documentation align with each other, as well as with the actual data being collected and what is displayed on the dashboard. 
If Publishers are not able to understand the documentation and/or the traffic analysis, they are not able to disclose in their privacy policies the data they collect and send to Data Recipients. The absence of such data disclosure infringes the \textit{Transparency principle}. 
This principle demands that Data Controller(s) must declare the categories of personal data they process, alongside %
purposes, recipients, risks and consequences for processing personal data (\cite[Art. 14(1)(d), Recital 39]{GDPR}~\cite[para. 40]{Case-Orange-Romania-SA-C-61-19}~\cite[para. 74]{CJEU-Planet49-19}).
Consequently, end-users cannot exercise their own individual rights (e.g., access, rectification) on the data they share, 
which infringes the \textit{Fairness principle} (Arts. 5(1)(a), 15-22 of the GDPR)~\cite{UKDPA-Data-Prot-Principles}.

\subsection{Automated study of all available Tags}
\label{sec:res-automated-study}

While automatically analyzing traffic generated by \tagstotal\ Tags, %
we identified multiple instances of data sharing with third-parties, addressing research question \textbf{RQ3}.

\subsubsection{Tags send requests to multiple third-parties}
\label{find:automated-multiple-companies}

\begin{figure}[tb]
    \centering
    \includegraphics[width=0.85\linewidth]{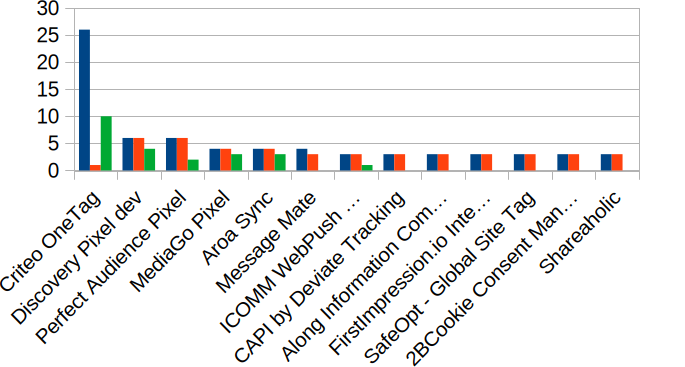}
    \caption{\textbf{Tags that contact three or more third-party companies}. Blue color highlights the  number of companies receiving a request initiated by the Tag; red  - the number of companies receiving the page URL; 
    green - the number of companies setting third-party cookies.}
    \label{fig:companies-cookies-per-tag}
\end{figure}

Our analysis of traffic generated by \tagstotal\ Tags shows that 352 (49\%) Tags send at least one request to a server\footnote{%
Tags are likely not sending as many requests as they would have sent if %
properly configured. See limitations in Section~\ref{sec:automated-study-limitations}.}. 
Moreover, 68 of such Tags communicate with multiple third-party domains, indicating that users’ data is potentially shared with other companies than the main Data Recipient associated to the Tag.  
Since some companies own multiple domains, we matched domains with companies (Section~\ref{sec:meth-automated-traffic-analysis}) and identified 49 Tags that send data to multiple companies. 
Moreover, out of these 49 Tags,  47 send the visited page URL in the HTTP ``Referer'' header to multiple companies and 9 Tags set third-party cookies. 

Figure~\ref{fig:companies-cookies-per-tag} focuses on the top 13 Tags (out of 49) that contact three or more companies, most of them also share the visited page URL and 6 Tags contact multiple companies that themselves set third-party cookies. 
For instance, Criteo OneTag sends requests to 26 third-party companies, 10 of which also set third-party cookies. 
This finding shows that Publishers %
need to be aware not only of the data shared by Tags with their own Data Recipients, but also about data shared and cookies set by other companies.
We therefore further analyze disclosures of the Data Recipients associated to such Tags. %

\subsubsection{Tags Providers do not always %
disclose the companies %
contacted by the Tag}
\label{find:not-declare-all-companies}

\begin{table}[t]
\caption{Tags sending requests to multiple third-parties.
``1p.'' signifies first-party and ``3p.'', third-party cookies. %
}
\begin{tabular}{p{2.1cm}p{1.2cm}|p{0.2cm}p{0.2cm}|p{0.3cm}p{0.3cm}|p{0.3cm}|p{0.2cm}p{0.5cm}}
\toprule
    \textbf{Tag}                                                      
    & \textbf{Provider}  
    & \begin{adjustbox}{angle=90} 
        {\# companies receiving requests }
        \end{adjustbox} 
    &\begin{adjustbox}{angle=90} 
        {\# undeclared companies} 
       \end{adjustbox} 
    & \begin{adjustbox}{angle=90} 
        {type of cookies}
        \end{adjustbox}
    & \begin{adjustbox}{angle=90} 
        {declared cookies set by the Tag} 
        \end{adjustbox}
    & \begin{adjustbox}{angle=90} 
        {\# other companies setting cookies}
        \end{adjustbox}
    & \begin{adjustbox}{angle=90} 
        {max. cookie duration (in days)}
        \end{adjustbox}\\
    \midrule
    
    Criteo One Tag                                              & Criteo SA         & 26    & 3 & 3p.& no  &  9  & 400  \\%no & 
    \rowcolor[HTML]{EFEFEF} Perfect Audience Pixel              & Sharpspring       & 6     & 4 & 3p.& no  &  1  & 400 \\%no & 
    Message Mate                                                & Owner Listens     & 4     & 1 & -   & -   &  -  &  -   \\%no & 
    \rowcolor[HTML]{EFEFEF} FirstImpression.io Integration Tag  & AppendAd          & 3     & 0 & -   & -   &  -  & -    \\ %
    Shareaholic                                                 & Shareaholic       & 3     & 2 & -   & -   &  -  &  -   \\%no & 
    \rowcolor[HTML]{EFEFEF} AWIN Conversion                     & Awin AG           & 2     & 1 & 1p. & no  &  -  & 365 \\%no & 
    Ternair Marketing Cloud Clicks Tracking                     & Ternair           & 2     & 0 & 1p. & no  &  -  & 400 \\  %
    \rowcolor[HTML]{EFEFEF} Amplitude Analytics Legacy          & Amplitude         & 2     & 1 & 1p. & no  &  -  & 365 \\%no & 
    Faslet's Size Me Up                                         & Faslet            & 2     & 1 & -   & -   &  -  & -    \\%no & 
    \rowcolor[HTML]{EFEFEF} Sprig Tag                           & Sprig             & 2     & 1 & -   & -   &  -  & -    \\%no & 
    Infoset                                                     & Infoset           & 2     & 1 & 1p. & yes &  -  & 400 \\ %
    \bottomrule
\end{tabular}
\label{tab:automatic-11-official-tags}

\end{table}

From the 49 Tags that send requests to multiple companies, we identified 11 official Tags (see details in Section~\ref{sec:official-tags}) and analyzed their legal documentation together with our legal co-author. %
Interestingly, we found only one Tag -- Ternair Marketing -- that explicitly mentions a Tag in its legal documentation.  
Table~\ref{tab:automatic-11-official-tags}
summarizes our findings. 
Only 3 out of 11 Tags disclose all companies contacted by the Tag. 
For instance, Perfect Audience Tag sends requests to 6 companies, and 4 of them are not disclosed, namely \textit{OpenX Technologies Inc.} (\texttt{openx.net}), \textit{Magnite, Inc.} (\texttt{rubiconproject.com}), \textit{Twitter, Inc.} (\texttt{twitter.com}) and \textit{Verizon Media} (\texttt{yahoo.com}).

\subsubsection{Tags set cookies without declaring them}
\label{find:not-declare-cookies}
We found that 6 out of 11 Tags set {first and} third-party cookies, but only %
Infoset Tag
specifies which cookies are set and mentions their purposes in the legal documentation\footnote{%
Such documentation is available in the supplemental materials~\cite{supplementary-materials}
}. %
Other Tags do not disclose specific cookies and their purposes, using phrasing such as 
``[t]hrough cookies placed on your browser or device, we may collect information''~\cite{SharpSpring-policy} or ``may store and access data on your device or browser, including by setting new cookies''~\cite{Criteo-policy}.

\subsubsection{When Tags contact multiple third-parties, some of them set cookies without disclosure}
\label{find:automated-fourth-cookies}

Additionally to setting their own cookies, Tags also send requests to other companies, some of which also set cookies. Out of the 49 such Tags, 
9 set third-party cookies. %
Out of 11 Official  Tags we have analyzed, 2 include requests to other companies that set their own cookies. For example, Criteo One Tag includes requests to 26 companies, and 9 of these companies set 21 different cookies\footnote{See Table~\ref{tab:criteo-cookies} in the Appendix listing all such cookies.}. %
This finding shows that Publishers need to be aware that Tags can create and share cookies not only with their own Data Recipients, but also with other companies.

\subsubsection{Compliance Consequences for Publishers due to lack of disclosures} 
When a Publisher installs a 
Tag that fails to  disclose all contacted third-party recipients, they are unaware of these entities, raising significant legal implications.
This lack of disclosure  hinders compliance 
with the \textit{Transparency principle} that requires  informing users about  data recipients and their rights
~\cite[Art. 14(1)(e), Recital 63]{GDPR},
\cite{CJEU-Planet49-19} %
\cite[para. 26]{Case-Osterrich-Post-C-154-21}.
It also obstructs %
 compliance with the \textit{Security principle} mandating %
 technical and organizational measures to prevent %
 unauthorized disclosure of personal data \cite[Arts. 5(1)(f), 32]{GDPR}. %

Furthermore, both Tags and multiple third-party companies set first- and third-party cookies without  being disclosed in the legal documentation (Sections~\ref{find:not-declare-cookies} and \ref{find:automated-fourth-cookies}). 
To determine whether a (first or third-party) cookie requires the legal basis of consent, \textit{purpose, duration and context} need to be analyzed~\cite{Foua-etal-20-IWPE, EDPB-4-12}. Since the purpose is never mentioned in the legal documentation, we analyze duration and context.
\emph{Duration} of at least one cookie set by each Tag is 365-400 days and in 2 out of 6 tags,  \emph{third-party cookies} are set (see Table~\ref{tab:automatic-11-official-tags}). 
Consent is required for such cookies.
Since the Publisher is unaware that certain tags/companies set such cookies, they cannot request the necessary consent from users for those cookies, which results in a violation of the \textit{Lawfulness principle}.

\subsubsection{Compliance consequences for Publishers regarding personal data}
\label{find:automated-personal-data}

Differently from the in-depth study, here  we do not extensively investigate all  Tags' documentation nor dashboard, and hence we do not analyze what data types are collected actively. 
However, when tags send requests to their own Data Recipients and to third-parties, the user's IP address and other browser- and OS-related data is \emph{passively sent}. 
Similar to our reasoning in in-depth study, %
we argue that data sent by Tags %
is potentially \textit{personal data} since Publishers and Data Recipients may have the means that are reasonably likely to identify end-users. They have access to the IP address and can combine it with additional identifiable data (Section~\ref{sec:in-depth-data-personal}).

\subsection{Comparative analysis}
\label{sec:comparative-analysis}
We compared the behavior of the 6 Tags analyzed in the in-depth study, against their behavior in our automated analysis (Section~\ref{sec:meth-comparison}).

Table~\ref{tab:method-comparison-6-tags} shows the types of requests made by the 6 tags during the two experiments.
We observe that the \textit{Google Tag} remained completely inactive in the automated study. %
The \textit{Hotjar Tracking Code} and \textit{Twitter Base Pixel} Tags still downloaded scripts as in the in-depth study but did not make any other request.
Only two Tags (i.e., \textit{LinkedIn Insight} and \textit{Quantcast Advertise}) behave the same regarding the number of scripts downloaded and requests sent.
In spite of our efforts in reasonably configuring the Tags (Section~\ref{sec:meth-automated}), the \textit{comScore Unified Digital Measurement} Tag could not be installed in our automated method.
\begin{table}[t]
\caption{Number of requests per type made by Tags in the in-depth and automated analysis.
Limitations of the automated analysis are highlighted in red.
}
\begin{tabular}{l|cc|cc|cc|cc|cc}
& 
\multicolumn{2}{p{0.9cm}}{\cellcolor[HTML]{000000}\centering \textcolor{white}{Google Tag}} &
\multicolumn{2}{p{0.9cm}}{\cellcolor[HTML]{000000}\centering \textcolor{white}{Twitter Base Pixel}} &
\multicolumn{2}{p{0.93cm}}{\cellcolor[HTML]{000000}\centering \textcolor{white}{Hotjar tracking Code}} & 
\multicolumn{2}{p{0.9cm}}{\cellcolor[HTML]{000000}\centering \textcolor{white}{Quantcast Advertise}} & 
\multicolumn{2}{p{0.9cm}}{\cellcolor[HTML]{000000}\centering \textcolor{white}{LinkedIn Insight}} 
\\

& 
\begin{adjustbox}{angle=90} in-depth \end{adjustbox} & \begin{adjustbox}{angle=90} automated \end{adjustbox}  &
\begin{adjustbox}{angle=90} in-depth \end{adjustbox} & \begin{adjustbox}{angle=90} automated \end{adjustbox}  &
\begin{adjustbox}{angle=90} in-depth \end{adjustbox} & \begin{adjustbox}{angle=90} automated \end{adjustbox}  &
\begin{adjustbox}{angle=90} in-depth \end{adjustbox} & \begin{adjustbox}{angle=90} automated \end{adjustbox}  &
\begin{adjustbox}{angle=90} in-depth \end{adjustbox} & \begin{adjustbox}{angle=90} automated \end{adjustbox}  \\ \hline                           

Script      & 1 & \textcolor{red}{0} & 1 & 1                  & 2 & \textcolor{red}{1} & 2 & 2 & 1 & 1  \\\rowcolor[HTML]{EFEFEF}
GET         & 0 & 0                  & 2 & \textcolor{red}{0} & 0 & 0                  & 1 & 1 & 1 & 1  \\
POST        & 1 & \textcolor{red}{0} & 0 & 0                  & 1 & \textcolor{red}{0} & 0 & 0 & 1 & 1  \\\rowcolor[HTML]{EFEFEF}
WebSocket   & 0 & 0                  & 0 & 0                  & 1 & \textcolor{red}{0} & 0 & 0 & 0 & 0  \\

\hline
\end{tabular}
\label{tab:method-comparison-6-tags}
\end{table}

Table~\ref{tab:method-comparison} further compares the data types collected during the two studies.
\textit{Google Tag} loads its script in the in-depth study and therefore sends multiple types of data only within the in-depth study (shown in red color).
\textit{Hotjar Tracking Code} loads only one script and therefore passive data is sent. However, it does not load the second script responsible for the WebSocket and the active collection of data.
\textit{Twitter Base Pixel} Tags only sends ``passive data'' (i.e., data sent as part of the HTTP communication, highlighted with the ``*'' in the ``network traffic'' columns in  Table~\ref{tab:method-comparison}) and cookies when downloading scripts, but do not send any other type of data.
The \textit{LinkedIn Insight} and \textit{Quantcast Advertise} Tags send the same data types. %

\subsection{Google Consent Mode v2}
\label{find:consent-mode}

We study Google Tag under Google Consent Mode v2 %
and analyze whether  %
it collects personal data lawfully, thus answering \textbf{RQ4}. 
We examine whether Google Tag disclosures %
correspond to the data collected during traffic analysis and explore the related  compliance issues.

Google consent mode v2 technical documentation mentions that %
Google Tag can run and will send \emph{``cookieless pings''} even when users decline consent.
It further declares that ``{[t]he data collected in the cookieless ping is used for behavioral and conversion modeling, to fill the gaps in your data}''~\cite{consent-mode-reference} and that it is used by Google AI to model the behavior of users who did not consent ``{at times when it is not possible to observe the path between ad interactions and conversions}''~\cite{about-consent-mode-modeling-snapshot-june2024}.

Regarding the data collected by such pings, the 
documentation says  that ``[c]ookieless pings, as part of regular HTTP/browser communication, may include the following information: user agent, screen resolution, IP address.''~\cite{consent-mode-snapshot-june2024}.
Moreover, the documentation says that when consent 
mode parameter \analyticsstorage\ is denied,  
``cookieless pings'' are sent,
but ``[n]o Analytics cookies are set, accessed, or read from the device''. 
Our experiment confirmed that
under the \emph{refuse all} condition (where all consent parameters are denied),
no cookies are indeed set or sent by the Google Tag. 
However, we have found several types of data sent by cookieless pings that lead to potential legal violations, described below. 

\subsubsection{Data is actively collected via ``cookieless pings'', but  declared in the documentation as passively collected}
\label{sec:consent-mode-cookieless-leaks}

\begin{table}[t]
    \centering
    \caption{Comparison of data types collected by the Tags in our two settings. Differences in data collected is shown in red. The presence of a star ``*'' indicates that this data type is sent as part of the HTTP(S) communication.}
    \includegraphics[width=\linewidth]{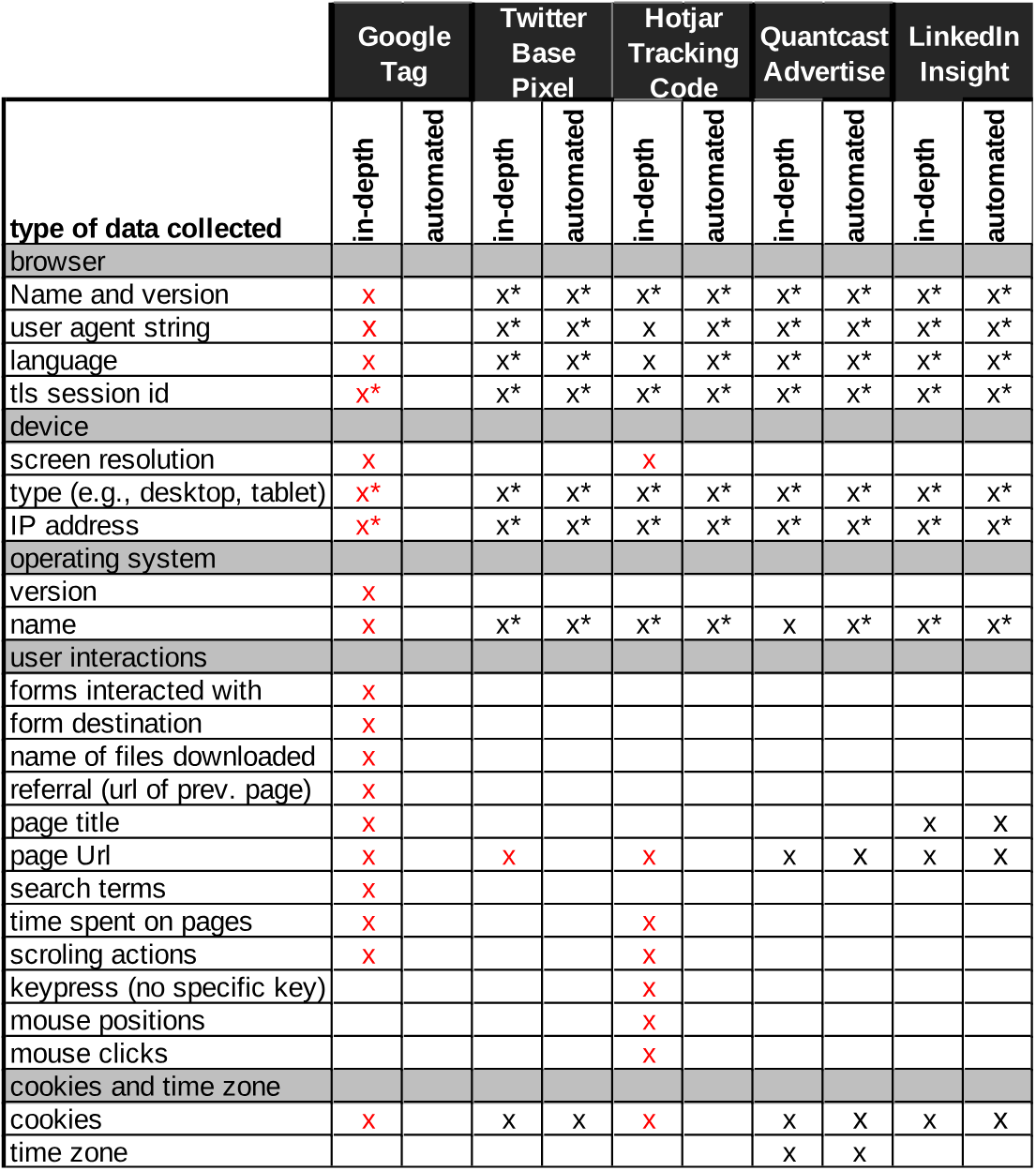}
    \label{tab:method-comparison}
\end{table}

In our experiment, we compared the URL parameters of all six requests\footnote{%
Example Table~\ref{tab:ga4_data_collected_without_consent}, full requests in supplemental materials\cite{supplementary-materials} %
} initiated by Google Tag under two consent choices : \emph{accept all} and \emph{refuse all} (Section~\ref{sec:meth-consent-mode}).
We found that the URL parameters under these two consent choices contain \emph{the same data}, meaning that the Google Tag \emph{actively collects}  %
 user language, %
 screen resolution, %
 computer architecture, %
 user agent string, %
 operating system and its version, %
 complete URL of the visited webpage,
 search keywords on the visited webpage%
 and sends it to the Google server.
Differently from Google's documentation that entails that  data is \emph{passively sent},
our results demonstrate that  \emph{data is actively collected} by Google Tag and added to the request in the URL parameters. 
Moreover, this data is always sent, regardless of the consent choices, since it is sent both under \emph{accept all} %
and  \emph{decline all} conditions. %

\subsubsection{Cookieless pings are declared to be anonymized upon user consent refusal}
\label{sec:consent-mode-anonymized}
Google claims that the data collected in cookieless pings is ``anonymized and non-identifiable Google Analytics events.''~\cite{consent-mode-reference}.
Our experiment confirms that
within cookieless pings, the \texttt{cid} parameter, instead of containing the value of the \texttt{\_ga} cookie (that is created when the user consents), contains a value randomly generated by the Google Tag on every web page visited and used instead.

\subsubsection{Compliance consequences for Publishers due to lack of disclosures}
\label{find:consent-personal-data}

Google Analytics server receives several types of user data even if users reject consent, as mentioned in Section~\ref{sec:consent-mode-cookieless-leaks} and %
Table~\ref{tab:ga4_data_collected_without_consent}.
In combination with these data types, it receives the IP address of the user which is actively collected by the Google Tag and added to the request in the URL parameters. 
Following the arguments presented in Section~\ref{sec:in-depth-data-personal}, we argue that the data collected by Google Analytics might be qualified as personal data. 
Moreover, when users reject consent for the (\adstorage and \analyticsstorage) variables, Google Analytics may collect and process personal data  \textit{without} a legal basis, thereby infringing the \emph{Lawfulness} principle.
We further claim the processing of personal data for behavioral and conversion modeling~\cite{about-consent-mode-modeling-snapshot-june2024} could possibly  infringe the \textit{Fairness principle} as users do not anticipate their data being collected after refusing consent.

Although Google claims that cookieless pings are anonymized and non-identifiable Google Analytics events~\cite{consent-mode-reference}, %
we argue users can still be indirectly identifiable by combining  the data mentioned in Section~\ref{sec:consent-mode-cookieless-leaks} with the user's IP address. 
Consequently, the claimed anonymization measure fails to meet the \textit{Data Protection by Default principle}, as it does not adequately  ensure that personal data is not processed by default~\cite[Art. 24(2)]{GDPR}.

Google Tag facilitates personal data transfers to Google LLC in the U.S. However, several regulators have deemed EU-U.S. data transfers to Google Analytics illegal because the required contractual, organizational, and technical measures are insufficient to prevent access by U.S. intelligence services~\cite{cnil-noyb-google-analytics-2020, dsb-noyb-google-analytics-2020}. Consequently, transferring unique identifiers, IP addresses, browser data, and metadata to Google LLC undermines the level of protection of personal data (Art. 44) and violates the \textit{Lawfulness and Security principles}. %

\subsection{Implications for Publishers}
\noindent\textbf{Data collected through Tags can be personal data.}
This paper argues that end-users can possibly be \textit{identifiable} since both Publishers and Data Recipients may have the means that are likely reasonably to be used to identify them through the combination of an IP address and several other data types, observed to be collected across the three studies (Sections~\ref{sec:in-depth-data-personal}, \ref{find:automated-personal-data}, \ref{find:consent-personal-data}). %

\noindent\textbf{Disclosures in the documentation of data collected  and data recipients  are not sufficient.}
Across %
our studies, we found that 
Data Recipients \textit{do not} declare in their documentation:
the %
Tag it refers to (Section~\ref{sec:meth-legal-tech-docs});
data recipients %
(Section~\ref{find:not-declare-all-companies}); 
cookies used by the Tag (Section~\ref{find:not-declare-cookies}); 
and declare data types in various places %
(Section~\ref{sec:in-depth-disclosure-issues}).

\noindent\textbf{Network traffic analysis is insufficient to detect derived and passive data collection.}
We found that Data Recipients store certain data which is either \textit{derived} %
from other types of data or \emph{passively sent} (Table~\ref{tab:in-depth-data}). 
For e.g., Google Tag technical documentation states that the city and country are derived from the user's IP address (Section~\ref{find:in-depth}).
Such example shows that Data Recipients collect and store more data types than what  directly appears in the network traffic.
Therefore, analyzing network traffic alone is insufficient to identify all the data types being collected.  Solely relying on disclosures is also inefficient.

\noindent\textbf{Legal implications for Publishers.}
 Data Recipients often place the burden onto the Publisher to understand their data collection practices. 
This implication might not be immediately clear to the Publisher %
because disclosures are often incomplete. 
The practice of insufficient disclosures across Tags makes it hard for the Publisher to comply with \textit{transparency} obligations. 
However, the lack of knowledge about the processing of personal data through Tags does not excuse the Publisher's \textit{accountability} obligations~\cite{Case-GoogleSpain}. 
Publishers must legitimize personal data processing with a legal basis, such as consent, when storing or reading identifiers like IP addresses, cookies, or other trackers from users' devices to comply with the \textit{Lawfulness principle}.

\subsection{Implications for Tag Providers}
\noindent \textbf{Tag Providers that are Data Recipients (official Tags).}
We found that certain %
official Tags send data to other companies without declaring it in the documentation (Section~\ref{find:automated-multiple-companies}); some set cookies without specifying them in documentation (Section~\ref{find:not-declare-cookies}), while others list data types collected in different places (Section~\ref{find:in-depth}), complicating Publisher's GDPR compliance. 
Attributing legal responsibility for these issues to Data Recipients is non-trivial. 
This is especially the case for Google Tag that leaks data even when users reject consent, resulting in the unlawful transfer of data to the US (Section~\ref{find:consent-personal-data}).
We contend that these actors cannot exonerate themselves of responsibility and shift it onto Publishers that include their Tags. 
Conversely, Publishers have a part of responsibility if they choose non-compliant Tags~\cite{Case-LiabilityController}. 
Hence, the responsibility of non-compliant Tags could potentially be shared between Data Recipients and Publishers, but we leave this discussion to %
future work.

\noindent \textbf{Tag Providers that are not Data Recipients (unofficial Tags).}
In the GTM configuration interface, each Template Gallery Tag is listed by name %
(e.g., "Facebook Pixel") along with its Tag Provider (e.g., Simo Ahava~\cite{ahava-facebook-pixel}).
However, when the Tag Provider is an %
unofficial entity and not a Data Recipient, it becomes harder to determine responsibility and  data collection disclosure obligations.

\section{Conclusion}

This work is the first to study the Tags within the Google Tag Manager, their behavior  and the data types they collect through three studies, including a first automated large-scale study of \tagstotal\ Tags. 
We discovered multiple hidden data leaks, unclear documentation, undisclosed third-parties and personal data
sharing without consent, 
and provided a legal analysis and derived compliance issues for our findings.
Our results demonstrate  compliance implications for Publishers,  the most critical being that disclosures in Tags documentation are incomplete and diverging. This fact introduces an important burden for Publishers  to understand the data collection practices of each specific Tag and all third-parties it includes, leaving Publishers struggling to  meet their legal obligations.

\section*{Acknowledgment}
This work has been supported by the ANR 22-PECY-0002 IPoP (Interdisciplinary Project on Privacy) project of the Cybersecurity PEPR, the TULIP project of the ANR MRSEI program 2023,  and the Inria International Chair funding. The authors would like to thank Michael Toth, former Inria PhD student, and Javiera Bermudez Alegria, from Universidad de Chile, who contributed to the initial work, and all the anonymous reviewers for helping us improve this paper.

\bibliographystyle{ieeetr}
\bibliography{urls,articles}

\renewcommand{\thesection}{\arabic{section}}
\appendix
\label{appendix}
\subsection{Data Availability}
\label{sec:data-availability}
\label{sec:artifacts}

Artifacts associated with this paper are available in a repository~\cite{supplementary-materials}.
We provide the network traces used for our in-depth and automatic analysis and for the consent mode case study.
We also provide the scripts to automate the collection the network traces %
with the explanation on how to configure and use them.%
The scripts to analyze the data produced are also provided.
For our in-depth study, we provide a copy of the documentation used to determine data types that the companies declare to collect.

\subsection{Cookies set by the Tags}

Tables~\ref{tab:criteo-cookies} and \ref{tab:perfect-cookies} list the cookies observed when studying Criteo OneTag and Perfect Audience Pixel within the automated study. 
 
\begin{table}[ht]
\caption{Cookies set when ``Criteo OneTag'' Tag was installed on our empty experimental website.}
\begin{tabular}{rcc}
\toprule
\textbf{cookie name}                & \textbf{cookie domain}          & \textbf{duration (days)} \\
\midrule
uid                        & .criteo.com            & 390             \\
ayl\_visitor               & .omnitagjs.com         & 30              \\
CMID                       & .casalemedia.com       & 365             \\
CMPS                       & .casalemedia.com       & 90              \\
CMPRO                      & .casalemedia.com       & 90              \\
receive-cookie-            &                        &                 \\
-deprecation               & .adnxs.com             & 400             \\
visitor-id                 & .media.net             & 365             \\
data-c-ts                  & .media.net             & 30              \\
data-c                     & .media.net             & 30              \\
mv\_tokens                 & exchange.mediavine.com & 14              \\
mv\_tokens\_eu-v1          & exchange.mediavine.com & 14              \\
am\_tokens                 & exchange.mediavine.com & 14              \\
am\_tokens\_eu-v1          & exchange.mediavine.com & 14              \\
criteo                     & exchange.mediavine.com & 14              \\
demdex                     & .demdex.net            & 180             \\
opt\_out                   & .postrelease.com       & 365             \\
dpm                        & .dpm.demdex.net        & 180             \\
C                          & .adform.net            & 31              \\
uid                        & .adform.net            & 60              \\
tvid                       & .tremorhub.com         & 365             \\
tv\_UICR                   & .tremorhub.com         & 30             \\
\bottomrule
\end{tabular}
\label{tab:criteo-cookies}
\end{table}

\begin{table}[ht]
\centering
\caption{Cookies set by the ``Perfect Audience Pixel'' Tag.}
\begin{tabular}{rcc}
\toprule
\textbf{cookie name}                & \textbf{cookie domain}          & \textbf{duration (days)}  \\
\midrule
pa\_uid             & .prfct.co     & 400             \\
pa\_twitter\_ts     & .prfct.co     & 400             \\
pa\_yahoo\_ts       & .prfct.co     & 400             \\
pa\_openx\_ts       & .prfct.co     & 400             \\
pa\_rubicon\_ts     & .prfct.co     & 400             \\
pa\_google\_ts      & .prfct.co     & 400             \\
personalization\_id & .twitter.com  & 400           \\
\bottomrule
\end{tabular}

\label{tab:perfect-cookies}
\end{table}

\subsection{Responsible disclosure notifications}
\label{appendix:responsible-disclosure-notification}

\subsubsection{Disclosure issue}

Dear Madam or Sir,

We are contacting you because you are listed in the imprint of the following service as the responsible party of \textit{company}.

During our research project, we identified practices regarding your Tag ``\textit{Tag}'' that might not fully comply with the EU data protection law.

We analysed the Google Tag Manager platform, alongside the Tags provided through it. We found that the Tag ``\textit{Tag}'' that you provide:
i) collects data types without fully disclosing them in your legal documentation; and/or
ii) sends data to third-parties without fully disclosing them in your legal documentation.
Providing such a Tag without proper disclosure of data  collected practices nor of the third-parties potentially infringes the law.

We believe there is a potential violation of Article 14(1)(d,e) and Recital 39 of the GDPR, as Data controller(s) must declare the categories of personal data they process, alongside purposes, recipients, risks and consequences for processing personal data. Such an infringement may be punishable by a fine in accordance with Article 83 GDPR.

The supervisory authorities can impose fines in case of violations of data protection law. These fines must be effective, proportionate and dissuasive in each individual case (cf. Art. 83 (1) GDPR). 
In the event of infringements of the principles governing processing pursuant to Article 14 (information obligations), fines of up to 20 million euros or, in the case of a company, up to 4\% of the total annual worldwide turnover achieved in the previous financial year, whichever is the higher, may be imposed (Art. 83 para. 5(b) GDPR).

For full details about our findings and potential legal violations, please refer to the sections 4.1 and 4.2 of our paper available here: https://arxiv.org/pdf/2312.08806
Should you need further information or have any other questions, please do not hesitate to contact us using the same email address. 

Sincerely, 

Cristiana Santos

Utrecht University School of Law

Email address: c.teixeirasantos@uu.nl

\subsubsection{Consent mode issue}

Dear Madam or Sir,

We are contacting you because you are listed in the imprint of the following service as the responsible party of Google.
During our research project, we identified practices regarding your Tag “Google Tag” that might not fully comply with the EU data protection law.

We analysed the Google Tag Manager platform, alongside the Tags provided through it. We found that the “Google Tag” that you provide collects data without user consent. Your Tag still collects personal data through cookieless pings without consent.
Collecting user data without consent potentially infringes the law. We believe there is a potential violation of Art. 6(1)(a) of the GDPR. Such an infringement may be punishable by a fine in accordance with Art. 83 GDPR.

Supervisory authorities can impose fines in the event of violations of data protection law. These fines must be effective, proportionate and dissuasive in each individual case (cf. Art. 83 (1) GDPR). In the event of infringements of the principles governing processing pursuant to Art. 5 GDPR and Art. 6 GDPR (lawfulness of processing), fines of up to 20 million euros or, in the case of a company, up to 4 %

For full details about our findings and potential legal violations, please refer to the section 4.3 of our paper available here: https://arxiv.org/pdf/2312.08806
Should you need further information or have any other questions, please do not hesitate to contact us using the same email address.

Sincerely, 

Cristiana Santos

Utrecht University School of Law

Email address: c.teixeirasantos@uu.nl

\begin{figure*}
    \centering
    \includegraphics[width=\linewidth]{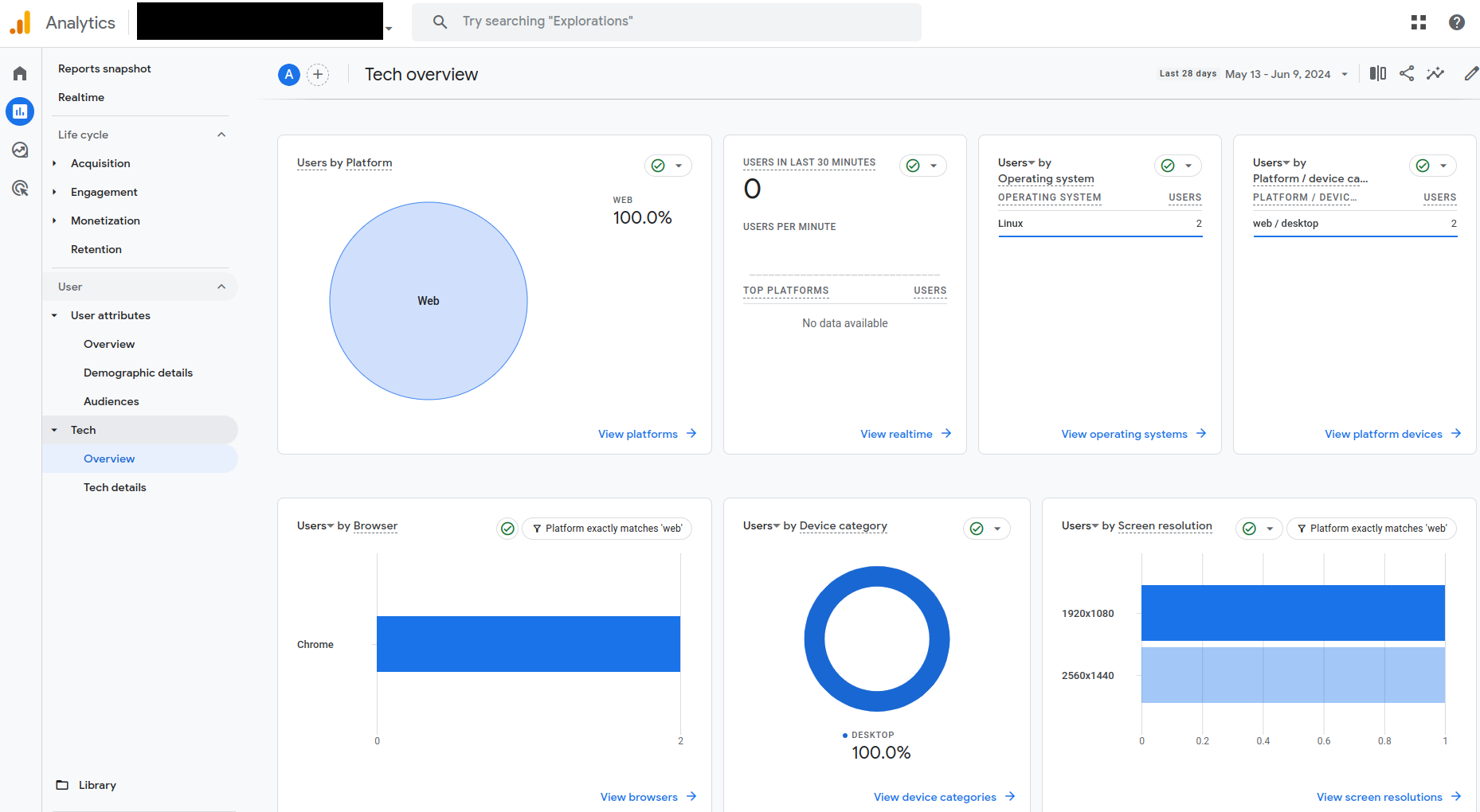}
    \caption{Dashboard of the Google Analytics service (Google Tag).}
    \label{fig:dashboard-example}
\end{figure*}

\begin{figure*}
    \centering
    \includegraphics[width=\linewidth]{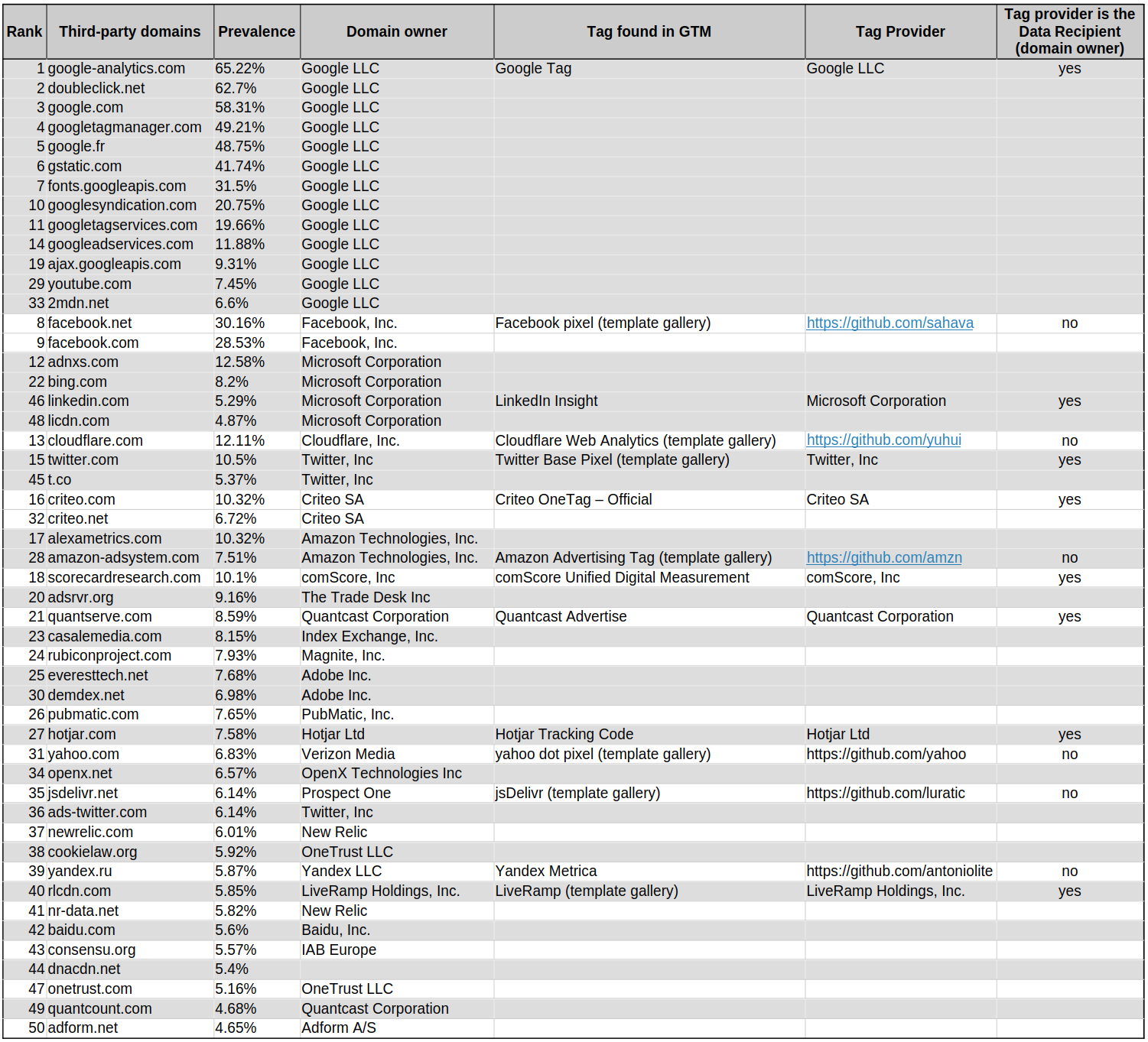}
    \caption{The Top 50 third party domains present on the top 1M websites (from~\cite{Foua-etal-22-PETs} with the company owning the domain and corresponding Tags we identified.}
    \label{fig:top-domains-tags-mapping}
\end{figure*}

\begin{table*}[ht]
    \centering
    \caption{
    Parameters found in the URL parameters or in the body of HTTP requests  to the \texttt{https://region1.google-analytics.com} 
    initiated by the Google Tag when studying Google consent mode v2 (Section~\ref{find:consent-mode}) under \emph{refuse all} and \emph{accept all}  conditions. The data that is identical in both requests is shown in \textbf{bold}. 
    A few additional parameters with no value are not shown in this table.
    }
  \begin{tabular}{p{1.9cm}|p{5.2cm}|p{5.2cm}|p{2cm}}
 \toprule
Query parameters & Condition \emph{refuse all}                                                                                                                     & Condition \emph{accept all}                                                                                                                      
& Type of data             
\\ 
\midrule
v               & \textbf{2} & 
\textbf{2}                                                                                     &                          \\
tid             & \textbf{G-XSKNFL9CWM}                                                                                                                    & \textbf{G-XSKNFL9CWM}                                                                                                                   &                          \\
gtm             & \textbf{45je43p0v9181346581za200}                                                                                                        & \textbf{45je43p0v9181346581za200}                                                                                                        &                          \\
\_p             &{  1711640211524}                                                                                             & { 1711640309156}                                                                                             &                          \\
gcs             & { G100 }                                                                                                     & { G111 }                                                                                                     &                          \\
gcd             & { 13p3p3p2p5 }                                                                                               & { 13t3t3t2t5 }                                                                                               &                          \\
npa             & { 1}                                                                                                         & { 0 }                                                                                                        &                          \\
dma\_cps        & { -}                                                                                                         & { sypham }                                                                                                   &                          \\
dma             & \textbf{1}                                                                                                                               & \textbf{1}                                                                             &                          \\
cid             & {  1337464789.17116}                                                                                         & { 1994365208.17116 }                                                                                         &                          \\
ul              &                        
\textbf{en-us}                                                                                                    &                       
\textbf{en-us}                                                                                                     & User language            \\
sr              &                       \textbf{2560x1440}                                                                                                &                        \textbf{2560x1440}                                                                                                & Screen resolution        \\
uaa             &                        
\textbf{x86}                                                                                                      &                        
\textbf{x86}                                                                                                      & Computer architecture    \\
uab             &                        
\textbf{64}                                                                                                       &                        
\textbf{64}                                                                                                       & Computer bitness         \\
uafvl           &                        \textbf{Chromium\%3B122.0.6261.94  ... }                                                                          &                        \textbf{Chromium\%3B122.0.6261.94 ... }                                                                           & User agent string        \\
uamb            & \textbf{0} &
\textbf{0}                                                                                    &                          \\
uap             &                        
\textbf{Linux}                                                                                                    &                        
\textbf{Linux}                                                                                                    & Operating system         \\
uapv            &                        
\textbf{6.6.17}                                               &                        
\textbf{6.6.17}                                                                                                   & Operating system version \\
uaw             & 
\textbf{0} &
\textbf{0} 
\\
pscdl           & { denied}                                                                                                    & { noapi}                                                                                                     &                          \\
\_s             & 
\textbf{2} & 
\textbf{2}  &                          
\\
sid             & { 1711640211}                                                                                                & { 1711640280}                                                                                                &                          \\
sct             & 
\textbf{1}                                                                                                                               & \textbf{1}                                                                                                                               &                          \\
seg             & { 0}                                                                                                         & { 1}                                                                                                         &                          \\
dr              & %
\textbf{https://ga4consent.example.com/consent 0000.html}

&  %
\textbf{https://ga4consent.example.com/consent 1111.html}
& Page url                 \\
dt              & \textbf{My experiment page  title}                                                                                               & \textbf{My experiment page  title}                                                         & Page title               \\
en              
& \textbf{view\_search\_results}                                                                                                           & 
\textbf{view\_search\_results}                                                                                                           &                          \\
tfd             & { 5620}                                                                                                      & { 5613 }                                                                                                     &                          \\
\_eu            & \textbf{AEA}                                                                                                                             & \textbf{AEA}                                                                                                                             &                          \\
\_et            & { 2 }                                                                                                        & { 1}                                                                                                         &                          \\
ep.search\_term & \textbf{my\%20search}                                                                                                                    & \textbf{my\%20search}                                                                                                                    & Search keyword           \\ \hline
\end{tabular}

    \label{tab:ga4_data_collected_without_consent}
\end{table*}

\end{document}